\documentclass[11pt,a4paper]{article}
\usepackage{amsmath}
\usepackage{authblk}%author affiliations
\usepackage{indentfirst}

\usepackage{amssymb}
\usepackage{braket}
\usepackage{xcolor}
 \usepackage[margin=0.75in]{geometry}
\usepackage{verbatim}
\usepackage{subcaption}
\usepackage{csquotes}

 \usepackage[normalem]{ulem}%strikethrough text
 \usepackage{cite}
 
 \usepackage{tikz}
\usepackage{lipsum}
\usepackage{float}

\newcommand{\Tr}{\rm{Tr}}
\usepackage{graphicx}

\title{Detection of Gravitational Waves using Parametric Resonance in Bose-Einstein Condensates}
\author[a,b,c]{Matthew P. G. Robbins \thanks{mrobbins@perimeterinstitute.ca}}
                                            \author[a,b,c]{Niayesh Afshordi \thanks{nafshordi@pitp.ca}}

                             \author[a,d]{Alan O. Jamison \thanks{alanj@uwaterloo.ca}}            \author[a,b,c,d]{Robert B. Mann \thanks{rbmann@uwaterloo.ca}}
                                          
                                            \affil[a]{Department of Physics and Astronomy, University of Waterloo,
                                            Waterloo ON, Canada, N2L 3G1}
                                           \affil[b]{Waterloo Centre for Astrophysics, University of Waterloo,
                                            Waterloo ON, Canada, N2L 3G1}
                                           \affil[c]{Perimeter Institute for Theoretical Physics,
                                            31 Caroline Street North, Waterloo ON, Canada, N2L 2Y5}
                                            \affil[d]{Institute for Quantum Computing, University of Waterloo,
                                            Waterloo ON, Canada, N2L 3G1}
\date{}

\begin{document}

\maketitle

\abstract{An interesting proposal for detecting gravitational waves involves quantum metrology of Bose-Einstein condensates (BECs). We consider a forced modulation of the BEC trap, whose frequency matches that of an incoming continuous gravitational wave. The trap modulation induces parametric resonance in the BEC, which in turn enhances sensitivity of the BEC to gravitational waves. We find that such a BEC detector could potentially be used to detect gravitational waves across several orders of magnitude in frequency, with the sensitivity depending on the speed of sound, size of the condensate, and frequency of the phonons. We outline a possible BEC experiment and discuss the current technological limitations. We also comment on the potential noise sources as well as what is necessary for such a detector to become feasible.
}

\section{Introduction}
The recent detection of gravitational waves is revolutionizing our understanding of the Universe. Though our current observations have only detected black holes and neutron stars with LIGO/Virgo \cite{Abbott2019},
it is expected that future gravitational wave detectors may also see phenomena such as supernovae, extreme-mass inspirals, and a stochastic gravitational wave background \cite{Amaro-Seoane2007,Christensen2019,Varun2019}.

One of the challenges of gravitational wave astronomy is that current detectors are sensitive to only a small range of frequencies. Constructing a multitude of detectors (both terrestrial and in space) is thus required to cover the entire gravitational wave spectrum \cite{Ballmer2015,Miller2019}. Much effort has been exerted into designing detectors for sub-kHz sources, such as the Laser Interferometer Space Antenna (LISA) (0.1 mHz-1 Hz) \cite{Amaro-Seoane2017}, the Einstein Telescope ($1-10,000$ Hz) \cite{Sathyaprakash2012,Maggiore2019} and pulsar timing arrays like the International Pulsar Timing Array and European Pulsar Timing Array ($\sim$nHz) \cite{Hobbs2010,Kramer2013}. Atom interferometers have also been proposed to detect gravitational waves in the millihertz to decahertz regimes \cite{Geiger2017,Gao2018,Coleman2018}. Though the Einstein Telescope is proposed to be sensitive up to 10 kHz, its optimal sensitivity will occur around the same frequency range as LIGO. Additional gravitational wave detector technologies have also been proposed, including a superfluid detector for continuous gravitational waves (0.1-1.5 kHz) \cite{Singh2017} and a satellite-based cold atom interferometer for mid-frequency gravitational waves (30 mHz-10 Hz) \cite{Graham2017}.

In the frequency regime  probed by LIGO, Virgo, and KAGRA (above 100 Hz and below 1 kHz), the most common sources are binary black holes, binary neutron stars, and black hole-neutron star mergers \cite{Abbott2018}. In the kHz regime, it is predicted that transient sources include lower-mass black holes and neutron star mergers  at frequencies outside the range that  LIGO can currently observe (up to several kHz) \cite{Maggiore2008,Andersson2003,Andersson2011}. Depending on the model, there could potentially even be primordial black holes in the kHz domain \cite{Bellido2017}. In the continuous regime, there may be magnetars (0.5-2 kHz) \cite{Stella2005}, neutron stars/pulars (tens-hundreds of Hz) \cite{Abbott2017,Covas2020}, and boson clouds (extending into the kHz regime and above) \cite{Riles2017}.

One intriguing proposal \cite{Sabin2014} suggests using a uniform (non-rotating) Bose-Einstein condensate (BEC) as a gravitational wave detector. By noting that an incoming gravitational wave is able to create phonons within the condensate and comparing the state of the phonons before and after the gravitational wave interacts with the BEC, it may be possible to gain knowledge about the amplitude of the gravitational wave.
Though a BEC is a quantum system, it is not possible to use the usual quantum mechanical method of doing measurements (projective-valued measurements, positive operator-valued measurements, etc.) to estimate the amplitude $\epsilon$ of the gravitational wave as the amplitude is not an operator observable (assuming the wave is classical). Instead, the quantum Fisher information $H_{\epsilon}$ can be used, which describes how much quantum information can be obtained from a single measurement of an arbitrary parameter in a quantum system. By doing $M$ measurements of the amplitude of the gravitational wave (for example,  using quantum dots or phonon evaporation \cite{Sabin2014}), the quantum Cramer-Rao bound states that the resultant uncertainty in the measurement of $\epsilon$ is $\braket{(\Delta\epsilon)^2}\geq\frac{1}{MH_\epsilon}$ \cite{Braustein1994}. Increasing either the number of measurements of the system or the quantum Fisher information obtained from each measurement therefore improves sensitivity to gravitational wave detection.

However, this proposal faces significant challenges.  We have previously pointed out that, while a BEC could in principle be used to detect gravitational waves, it is necessary to squeeze the initial state of the BEC phonons well beyond current experimental capabilities \cite{Robbins2019}. Furthermore, the natural decay of the BEC by Beliaev damping limits the sensitivity, especially at the higher frequencies of interest.  A somewhat stronger critique 
 \cite{Schutzhold2018}  argued that non-uniform condensates are required, as otherwise the phonons produced are not within the realm of detectability, though it should be noted that the critique assumed
 a condensate without any boundary conditions, whereas the original proposal \cite{Sabin2014} considered a hard-wall condensate. 
 
We investigate here the possibility of improving the sensitivity of a BEC detector by modulating the trap size to induce a parametric resonance with the BEC phonons, which then boosts their sensitivity to gravitational waves. In contrast to previous studies \cite{Sabin2014,Robbins2019}, we find that parametric resonance causes the BEC to be more sensitive to gravitational waves at lower frequencies than at higher frequencies, though
detection of  higher frequency gravitational waves may still be possible. The optimum sensitivity depends upon the parameters of the condensate such as its length, speed of sound and phonon frequency.

The outline of the subsequent sections is as follows: In Section \ref{sec: BECs in curved spacetime}, we review the theory behind BECs in curved spacetime and solve the equations of motion of the phonons as well as the Bogoliubov coefficients. We also discuss the effect of parametric resonance and non-linearities in our model. In Section \ref{sec: Estimation}, we then show how quantum Fisher information is used to determine the sensitivity to gravitational waves. Next, in Section \ref{sec: damping}, we discuss how Beliaev damping will limit the sensitivity of the BEC to gravitational waves at high frequencies. In Section \ref{sec: experimental}, we discuss the possible sources that a BEC could observe as well as consider the experimental feasibility of the proposal, comment on expected noise sources, and discuss what may be necessary to make this proposal a reality. Section \ref{sec: Conclusion} summarizes our conclusions and presents possible directions for future work.

\section{Bose-Einstein condensates in curved spacetimes}
\label{sec: BECs in curved spacetime}

\subsection{Equation of Motion}
\label{sec: EOM}

We first derive the equations of motion and Bogoliubov coefficients for BEC phonons in a curved spacetime. Note that such derivations have also been done in \cite{Sabin2014,Nicolis2014,Fagnocchi2010}. The Lagrangian for a BEC in a curved spacetime is
\begin{align}
\mathcal{L}=g^{\mu\nu}\partial_{\mu}\phi\partial_{\nu}\phi^{\dagger}-M^2(x^i,t)|\phi|^2-U(|\phi|^2) ,
\label{eq: L}
\end{align}
where  $g_{\mu\nu}$ is the metric, $\phi$ is the field, and $U(|\phi|^2)=\lambda|\phi|^4>0$ is the interaction potential. The quantity $M^2(x^i,t)=\frac{m^2c^2}{\hbar^2}+V(x^i,t)$ describes a time-dependent effective mass of the atoms in the condensate, where we take
\begin{align}
V(x^i,t)=V_0\prod_{i=1}^3\Theta\left[\frac{x^i}{\alpha(t)}\left(\frac{x^i}{\alpha(t)}-L\right)\right]
\end{align}
and $V_0$ is a constant. This simple, separable effective potential models the effect of trap boundaries oscillating at frequency $\Omega_B$, which in turn act to cause the length of the condensate to oscillate. 
The quantity $\alpha(t)=1+a\sin\Omega_B t$ describes the fluctuations of the walls of the trap with $a=\frac{\delta L}{L}\ll1$ the variation of the box potential length.

Given the definition $\phi=\hat{\phi}e^{i\chi}$ with real $\hat{\phi}$ and $\chi$, and assuming a slowly varying amplitude, i.e. that $(\partial \hat{\phi})^2 \ll \hat{\phi}^2(\partial \chi)^2$, we can extremize $\mathcal{L}$ with respect to $\hat{\phi}$ to find: 
\begin{align}
\hat{\phi}^2=\frac{1}{2\lambda}\left[\partial^{\mu}\chi\partial_{\mu}\chi-m^2-V(x^i,t)\right]\ , 
\label{eq: phi2}
\end{align}
and so 
\begin{align}
\mathcal{L}=\frac{1}{2\lambda}\left[g^{\mu\nu}\partial_{\mu}\chi\partial_{\nu}\chi-\frac{m^2c^2}{\hbar^2}-V(x^i,t)\right]^2\ .
\label{eq: L2}
\end{align}
To study the effects of a gravitational wave, we introduce the coordinate system   $\bar{x}^i=\frac{x^i}{\alpha}$ and $ct=x^0=\bar{x}^0$. In the $\bar{x}^{\mu}$ coordinate system, the line element becomes 
\begin{align}
ds^2=\left(g_{00}+g_{ij}\bar{x}^i\bar{x}^j\dot\alpha^2\right)(d\bar{x}^0)^2+2g_{ij}\bar{x}^j\alpha\dot\alpha d\bar{x}^0d\bar{x}^i+g_{ij}\alpha^2d\bar{x}^id\bar{x}^j\ ,
\end{align}
where we are using a $(+,-,-,-)$ signature.
Let $\chi=\frac{mct}{\hbar}+\pi(x^\mu)$, where $\pi(x^\mu)\in\Re$ is a pseudo-Goldstone boson representing the acoustic perturbation (phonon) field of the condensate. These phonons can be represented as fluctuations in the phase \cite{Schutzhold2007}.
Setting $g_{ij}=\eta_{ij}+h_{ij}$,  with $[h_{ij}]\ll1$, the action of the system becomes
\begin{align}
S=\frac{1}{2\lambda}\int\left[A\dot\pi^2/c^2+C^i\bar\partial_i\pi+D^i\dot\pi\bar\partial_i\pi/c+E^{ij}\bar\partial_i\pi\bar\partial_j\pi\right]\sqrt{-\bar g}d^4x\ ,
\label{eq: S}
\end{align}
where $h_{ij}$ describes the distortion of spacetime due to the gravitational wave and
\begin{align*}
A/c^2&=\frac{1}{c^2}\left[-2 V-2\frac{m^2c^2}{\hbar^2}+\frac{6\kappa^2}{\hbar^2 c^2}\right]+\mathcal{O}\left(\frac{\Omega_B^2L^2}{c^2}\right)\ ,\\
C_i/c&=\frac{a\Omega_B}{c}\left[-\frac{4cm^2\kappa}{\hbar^3}+\frac{4\kappa^3}{\hbar^3 c^3}-\frac{4V\kappa}{\hbar c}\right]g_{ij}\bar{x}^j\cos\Omega_B t +\mathcal{O}\left(\frac{\Omega_B^2L^2}{c^2}\right)\ ,\\
D_i/c&=\frac{a\Omega_B}{c}\left[-4V-\frac{4c^2m^2}{\hbar^2}+\frac{12\kappa^2}{\hbar^2 c^2}\right]g_{ij}\bar{x}^j\cos\Omega_B t+\mathcal{O}\left(\frac{\Omega_B^2L^2}{c^2}\right)\ ,\\
E_{ij}/c^2&=\frac{1}{c^2}\left[-2V-\frac{2m^2 c^2}{\hbar^2}+\frac{2\kappa^2}{\hbar^2 c^2}\right]g_{ij}\left(1+2a\sin\Omega_B t\right)+\mathcal{O}\left(\frac{\Omega_B^2L^2}{c^2}\right)\ ,
\end{align*}
as $x^i\lesssim L$. We denote differentiating with respect to $\bar{x}^i$ as $\bar\partial_i$ and assume that $\pi$ vanishes at the boundary.

As $V$ consists of Heaviside functions, we can treat   the cases of $V=0$ and $V=V_0$ separately. The only difference in what follows will be the value of the speed of sound. Therefore, the equation of motion to leading order in $a$ is
\begin{equation}
\begin{aligned}
\ddot\pi+c_s^2g^{ij}(1+2a\sin\Omega_Bt)\bar{\partial}_i\bar{\partial}_j\pi=0\ ,
\label{eq: EOM}
\end{aligned}
\end{equation}
where
\begin{align}
c_s^2=\frac{-V_0-\frac{m^2c^2}{\hbar^2}+\frac{\kappa^2}{\hbar^2c^2}}{-V_0-\frac{m^2c^2}{\hbar^2}+\frac{3\kappa^2}{\hbar^2c^2}}c^2
\end{align}
is the speed of sound, assuming a linear dispersion relation for the phonons.  Setting $\pi=e^{i\vec{k}\cdot \bar{x}}\psi(t)$, we have
\begin{equation}
\begin{aligned}
\ddot\psi+c_s^2g_{ij}(1+2a\sin\Omega_Bt)k^ik^j\psi=0\ ,
\end{aligned}
\label{psidd}
\end{equation}
where we have neglected terms $\mathcal{O}\left(\frac{\Omega_B^2L^2}{c^2}\right)$ and higher. A derivation of the dispersion relation for the BEC for  $V_0\neq 0$ is
similar to that of the  $V=0$ case \cite{Robbins2019}, with $m$ becoming a time-dependent effective mass, such that $\frac{m^2c^2}{\hbar^2}\to \frac{m^2c^2}{\hbar^2}+V_0$. In the limit of $V_0 > 2 m \mu /\hbar^2$, the phonons cannot propagate outside the walls, and thus we can approximately use Dirichlet boundary conditions. 

By working in the TT-gauge and assuming no cross-polarization for simplicity,
\begin{align}
h_{\mu\nu}=\begin{pmatrix}
0 & 0 & 0 & 0\\
0 & h_+(t) & 0 & 0\\
0 & 0 & -h_+(t) & 0\\
0 & 0 & 0 & 0
\end{pmatrix}\ .
\end{align}
Taking $h_+(t)=\epsilon\sin\Omega t$, where $\epsilon$ is the amplitude of the continuous gravitational wave and $\Omega$ is its frequency, the equation of motion \eqref{psidd} becomes
\begin{align}
\ddot\psi + \omega^2(1+2a\sin\Omega_Bt)(1+\tilde\epsilon\sin\Omega t)\psi=0\ ,
\label{eq: EOM2}
\end{align}
where $\tilde{\epsilon}=\frac{(k_x^2-k_y^2)}{|k|^2}\epsilon$.

Let $\omega_1(t)^2=\omega^2(1+\tilde\epsilon\sin\Omega t+2a\sin\Omega_Bt+2a\tilde\epsilon\sin\Omega t\sin\Omega_Bt)$. As we want to consider the gravitational wave on resonance with the magnetic field, we will take $\Omega_B=\Omega$. Expanding to first-order in $\epsilon$ and second-order in $a$ yields $\omega_1(t)\approx\omega+ a \omega  \sin (\Omega t)$
   where we have neglected the $\mathcal{O}(a\tilde\epsilon)$ terms as $a\tilde\epsilon\ll a$. We will see that it is the presence of the $\mathcal{O}(a)$ terms that induce parametric resonance within the condensate.

To solve equation (\ref{eq: EOM2}) in this approximation, let us assume that we can decompose the solutions as \cite{Braden2010}
\begin{align}
\pi&=\alpha(t)\frac{e^{-i\int\omega_1(t')dt'}}{\sqrt{2\omega_1(t)}}+\beta(t)\frac{e^{i\int\omega_1(t')dt'}}{\sqrt{2\omega_1(t)}}\ ,\\
\dot\pi&=-i\alpha(t)\sqrt{\frac{\omega_1(t)}{2}}e^{-i\int\omega_1(t')dt'}+i\beta(t)\sqrt{\frac{\omega_1(t)}{2}}e^{i\int\omega_1(t')dt'}\ ,
\end{align}
where $\alpha(t)$ and $\beta(t)$ are time-dependent Bogoliubov coefficients that satisfy the coupled differential equations
\begin{equation}
\begin{aligned}
\dot\alpha&=\frac{\dot\omega_1}{2\omega_1}e^{2i\int\omega_1(t')dt'}\beta(t) \qquad 
\dot\beta&=\frac{\dot\omega_1}{2\omega_1}e^{-2i\int\omega_1(t')dt'}\alpha(t) 
\label{eq: alpha, beta}
\end{aligned}
\end{equation}

We will solve these coupled equations using the method outlined in \cite{Arnold1989}.
Setting $\hat\alpha=e^{-i\bar\omega t}\alpha$ and $\hat\beta=e^{i\bar\omega t}\beta$, where $\bar\omega=\frac{1}{T}\int_0^T\omega_1(t)dt$,  we obtain from \eqref{eq: alpha, beta},
\begin{align}
\frac{d}{dt}\begin{pmatrix} \hat\alpha\\ \hat\beta \end{pmatrix}=
\begin{pmatrix}
-i\bar\omega & \frac{\dot\omega_1}{2\omega_1}e^{2i\int\delta\omega(t')dt'}\\
\frac{\dot\omega_1}{2\omega_1}e^{-2i\int\delta\omega(t')dt'} &i\bar\omega
\end{pmatrix}
\begin{pmatrix}
\hat\alpha \\
\hat\beta
\end{pmatrix}\ ,
\label{eq: hat}
\end{align}
where  $\delta\omega(t)=\omega_1(t)-\bar\omega$.
Let $\hat\alpha=\hat\alpha_0+b\hat\alpha_1+b^2\hat\alpha_2$ and $\hat\beta=\hat\beta_0+b\hat\beta_1+b^2\hat\beta_2$, where $b=2a+\epsilon\ll1$. Note that, up to second order in $b$, $\bar\omega=\omega\left(1-\frac{b^2}{16}\right)$.

Substitution into equation (\ref{eq: hat}), solving order-by-order, and restricting ourselves to solving the system after each period, we can define:
\begin{align}
\begin{pmatrix}
\hat\alpha(T)\\
\hat\beta(T)
\end{pmatrix}
=M\begin{pmatrix}
\hat\alpha(0)\\
\hat\beta(0)
\end{pmatrix}=\begin{pmatrix}
M_{11} & M_{12}\\
M_{12}^* & M_{22}
\end{pmatrix}\begin{pmatrix}
\hat\alpha(0)\\
\hat\beta(0)
\end{pmatrix}\ ,
\end{align}
where after a a period $T=2\pi/\Omega$, we have
\begin{align}
M_{11}&=\frac{e^{-2 i \pi  q} \left(4+q^2 \left(b^2 \left(-\left(-8 i \pi  q^3+2 i \pi  q+e^{4 i
   \pi  q}-1\right)\right)+64 q^2-32\right)\right)}{4 \left(1-4 q^2\right)^2}\ ,\\
M_{12}&=\frac{b q \left(-3 i b q+4 q^2-4\right) \sin (2 \pi  q)}{4 \left(4 q^4-5 q^2+1\right)}\ ,\\
M_{22}&=\frac{e^{-2 i \pi  q} \left(-b^2 q^2+e^{4 i \pi  q} \left(4+q^2 \left(b^2 \left(-8 i \pi
    q^3+2 i \pi  q+1\right)+64 q^2-32\right)\right)\right)}{4 \left(1-4 q^2\right)^2}\ .
\end{align}
We have assumed that $\hat\alpha_1(0)=\hat\beta_1(0)=\hat\alpha_2(0)=\hat\beta_2(0)=0$. We see that the 1st order resonance (i.e. to 2nd order in $b$) occurs when $q\approx\frac{1}{2}$ (see Figure \ref{fig: Mathieu}).

After $N$ periods of the gravitational wave oscillations (or equivalently after $N$ periods of the trap modulation), we have
\begin{align}
\begin{pmatrix}
\hat\alpha(NT)\\
\hat\beta(NT)
\end{pmatrix}
=
\begin{pmatrix}
e^{-2\pi iqN}\alpha(NT)\\
e^{2\pi iqN}\beta(NT)
\end{pmatrix}
=M^N\begin{pmatrix}
\hat\alpha(0)\\
\hat\beta(0)
\end{pmatrix}\ .
\end{align}

Diagonalizing $M$ and writing $\begin{pmatrix}
\hat\alpha(0)\\
\hat\beta(0)
\end{pmatrix}=k_1\lambda_1\vec{x}_1+k_2\lambda_2\vec{x}_2$, where $\lambda_i,\vec{x}_i$ are the eigenvalues/eigenvectors of $M$ and $k_i$ are coefficients related to the eigenvalues, we see

\begin{align}
\begin{pmatrix}
\hat\alpha(NT)\\
\hat\beta(NT)
\end{pmatrix}=k_1\lambda_1^N\vec{x}_1+k_2\lambda_2^N\vec{x}_2\ .
\end{align}

Let $\lambda_2$ correspond to the eigenvalue whose magnitude is less than one around resonance. We can neglect this term because it will become negligible after many periods. Therefore, the only relevant quantities are

\begin{align}
    k_1&=-\frac{b \left(128 \zeta ^3-64 i \zeta ^2+3 \zeta +i\right)+128 \zeta ^2-2}{4 \left(1-64
   \zeta ^2\right)^{3/2}}\\
    \lambda_1&=\pi  b^2 \left(\frac{5 \sqrt{1-64 \zeta ^2} \zeta }{512 \zeta ^2-8}+\pi  \left(2 \zeta
   ^2-\frac{1}{32}\right)\right)-\frac{1}{4} \pi  b \sqrt{1-64 \zeta ^2}-1
\end{align}
where $\zeta\equiv(q-\frac{1}{2})/b$, and
\begin{align}
    \vec{x}_1=\begin{pmatrix}
    x_0+bx_1+b^2x_2\\
    1
    \end{pmatrix}
\end{align}
with
\begin{align}
    x_0&=\sqrt{1-64 \zeta ^2}-8 i \zeta\\
    x_1&=8 i \zeta ^2+\frac{\zeta  (3+64 \zeta  (2 \zeta -i))+i}{2 \sqrt{1-64 \zeta ^2}}+4 \zeta
   +\frac{3 i}{16}\\
    x_2&=-16 i \zeta ^3+\frac{28 \zeta ^2}{3}-\frac{1}{12} i \pi ^2 \zeta +\frac{37 i \zeta
   }{16}-\frac{3}{32}\\
   &+\frac{-219+128 \zeta  \left(4 \zeta  \left(2 \pi ^2 \left(64 \zeta ^2-1\right)+16 \zeta 
   (4 \zeta  (-63+32 \zeta  (12 \zeta +7 i))-43 i)+75\right)+29 i\right)}{1536
   \left(1-64 \zeta ^2\right)^{3/2}}
\end{align}
Let us further define $\alpha_{2a+\tilde\epsilon}=\alpha_{2a}+\tilde\epsilon\alpha_{\tilde\epsilon}$ and  $\beta_{2a+\tilde\epsilon}=\beta_{2a}+\tilde\epsilon\beta_{\tilde\epsilon}$. The goal is to calculate the Bogoliubov coefficients $\alpha_{\tilde\epsilon}$ and $\beta_{\tilde\epsilon}$, which correspond to the effect solely due to the of the gravitational wave interacting with the trap modulation. Using the definition of the Bogoliubov coefficients, we can write $\hat a_0=\alpha_{2a+\tilde\epsilon}\hat{a}_{2a+\tilde\epsilon}+\beta^*_{2a+\tilde\epsilon}\hat{a}_{2a+\tilde\epsilon}^\dagger$, where $\hat a_0$ is the the annihilation operation in the mode decomposition with no gravitational wave or modulation and $\hat{a}_{2a+\tilde\epsilon}$ is the annihilation operator in the mode decomposition containing both a gravitational wave and magnetic field. Thus $\hat{a}_{2a+\tilde\epsilon}=\alpha^*_{2a+\tilde\epsilon}\hat{a}_0-\beta_{2a+\tilde\epsilon}^*\hat{a}_0^{\dagger}$. We can also write $\hat{a}_{2a+\tilde\epsilon}=\alpha_{rel}\hat{a}_{2a}-\beta_{rel}^*\hat{a}_{2a}^\dagger$. Therefore,
\begin{align}
\hat{a}_{2a+\tilde\epsilon}&=\alpha^*_{2a+\tilde\epsilon}\left(\alpha_{2a}\hat{a}_{2a}+\beta^*_{2a}\hat{a}_{2a}^\dagger\right)-\beta_{2a+\tilde\epsilon}^*\left(\alpha_{2a}^*\hat{a}_{2a}^\dagger+\beta_{2a}\hat{a}_{2a}\right)\\
&=\left(\alpha^*_{2a+\tilde\epsilon}\alpha_{2a}-\beta_{2a+\tilde\epsilon}^*\beta_{2a}\right)\hat{a}_{2a}+\left(\alpha^*_{2a+\tilde\epsilon}\beta^*_{2a}-\beta_{2a+\tilde\epsilon}^*\alpha_{2a}^*\right)\hat{a}_{2a}^\dagger \ .
\end{align}
With $\alpha_{2a+\tilde\epsilon}=\alpha_{2a}+\tilde\epsilon\alpha_{\tilde\epsilon}$ and $\beta_{2a+\tilde\epsilon}=\beta_{2a}+\tilde\epsilon\beta_{\tilde\epsilon}$, we see
\begin{align}\
\label{eq: alpha rel}
\alpha_{rel}^*&=1+\tilde\epsilon\left(\alpha^*_{\tilde\epsilon}\alpha_{2a}-\beta_{\tilde\epsilon}^*\beta_{2a}\right)\ ,\\
\label{eq: beta rel}
\beta_{rel}^*&=-\tilde\epsilon\left(\alpha^*_{\tilde\epsilon}\beta^*_{2a}-\beta_{\tilde\epsilon}^*\alpha_{2a}^*\right)\ .
\end{align}

\subsubsection{Resonance}
\label{sec: Resonance}
\begin{figure}[b!]
\centering
\begin{subfigure}{0.49\textwidth}
\centering
\includegraphics[width=\columnwidth]{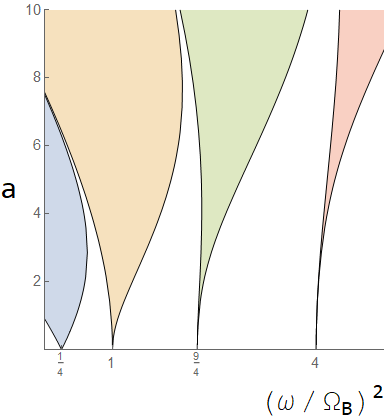}
\caption{Shaded areas show the locations of instability bands in $(a,\omega)$ plane, associated with resonances of equation \eqref{eq: EOM2} 
with $\Omega_B=\Omega$. By lying within these regions for a time $t$, it is possible to improve the sensitivity to gravitational waves. Note that maximum value of $\omega$ will be constrained by the chemical potential of the condensate. }
\end{subfigure}
\begin{subfigure}{0.49\textwidth}
\includegraphics[width=\columnwidth]{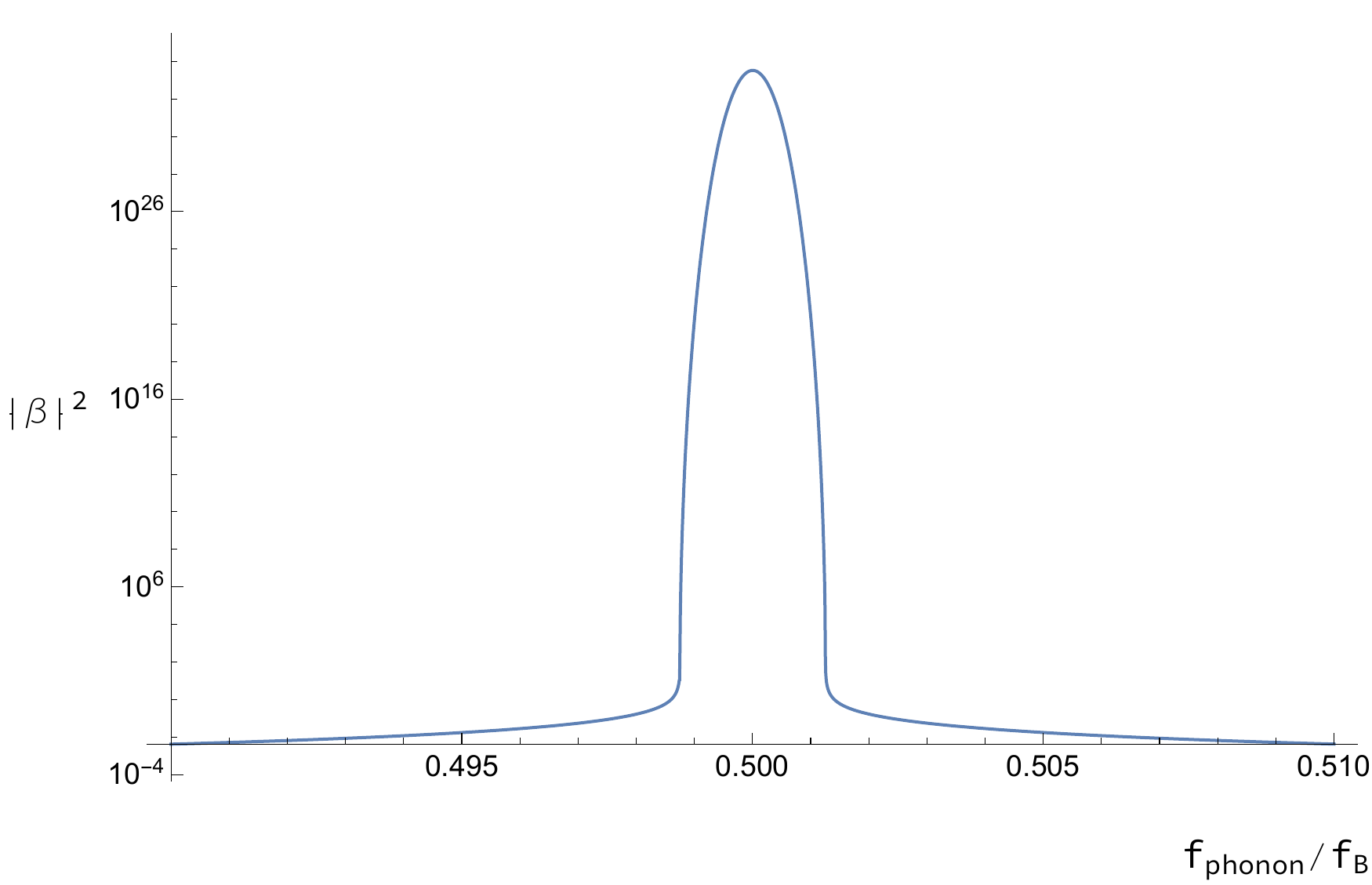}
\caption{The first resonance of our undamped system, where for illustrative purposes, we set $N=5\times10^3$ and $a=0.005$ Note that $|\beta|^2$ is maximized around $q\approx\frac{1}{2}$, before there is a sharp cut-off around $q=0.499$ and $q=0.501$. This corresponds to moving outside the instability band.}
\end{subfigure}
\caption{}
\label{fig: Mathieu}
\end{figure}

Equation \eqref{eq: EOM2}   is a perturbed form of the Mathieu differential equation (as $a\tilde\epsilon$ is a sub-dominant term), which implies that our system includes parametric resonance effects \cite{DLMF, Morse1953}. 

In Figure \ref{fig: Mathieu}a, we show the locations of the first four resonances of the system, where the shading illustrates the instability bands associated with each individual resonance in the Mathieu equation (Figure \ref{fig: Mathieu}a) \cite{Mclachlan1947}. For a system lying within these bands (indicated by the shaded regions), evolution is unstable and will become non-perturbative after a certain amount of time, with the amount of time before non-perturbative effects occurs depending on the precise location within these regions. In Figure \ref{fig: Mathieu}b, we see that if $q=\frac{\omega}{\Omega}$ is too far off the resonance, the instability band is exited, which limits production of particles and nonlinearities.

By considering phonons with frequencies that that are within the instability bands (locations of parametric resonance) for a time $t$, we can enhance the sensitivity of the BEC to gravitational waves. In Section \ref{sec: Non-linearities}, we address the maximum time that we can lie within the regions of instability before the system becomes non-perturbative in $a$.

\subsubsection{Non-linearities}
\label{sec: Non-linearities}

For $\kappa\approx m$, similar to \cite{Robbins2019}, we have the constraint of $\dot\pi\ll \mu = mc_s^2$. By squaring both sides (as $\braket{\dot\pi}=0$) and writing $\dot\pi$ as a sum of creation and annihilation operators, we find that this is equivalent to $\sum_{\vec k}|\beta_{2a,\vec{k}}|^2\hbar\omega_{\vec{k}}\ll n m c^2_s L^3$, where $L^3$ is the volume of the condensate (taking the BEC to be cubic with sides of length $L$), $n$ is the number density, $m$ is the mass of the atoms, and $|\beta_{2a,\vec{k}}|^2$ is the number of phonons present in each mode $\vec k$ \footnote{Accounting for the effect of the gravitational wave on this energy condition is negligible.}. 

Rewriting $|\beta_{2a,\vec{k}}|^2$ in terms of $q=\frac{\omega}{\Omega}$ and assuming that the BEC is constructed such that only a single phonon mode dominates the resonance instability  (i.e., the most unstable mode), we note that at each value of $q$, there is a maximum number of trap oscillations $N(q)$ satisfying $\frac{|\beta_{2a,q}|^2\hbar\omega_{q}}{\rho L^3}\lessapprox0.05$, where we relabel $|\beta_{2a,\vec{k}}|^2$ as $|\beta_{2a,q}|^2$ to explicitly indicate the dependence on $q$. Physically, $N(q)$ describes the maximum time that a gravitational wave could be observed by the condensate before the condensate becomes non-linear (because, by assumption, we take the frequency of the trap oscillation to be the same as the period of the gravitational wave). Noting that $|\beta_{2a,q}|^2=|k_1|^2|\lambda_1|^{2N}$, we see that

\begin{align}
    N\leq \frac{\log\left[\frac{0.05mnc_S^2L^3}{\pi\hbar qf_{GW}|k_1|^2}\right]}{2\log[|\lambda_1|]}\ .
\end{align}

\begin{figure}[t!]
    \begin{subfigure}{\textwidth}
\centering
\includegraphics[scale=0.7]{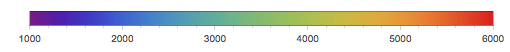}
\caption*{Maximum number of gravitational wave periods detectable by a BEC.}
\end{subfigure}\vspace{0.5cm}
\begin{subfigure}{.3\textwidth}
  \includegraphics[width=\linewidth]{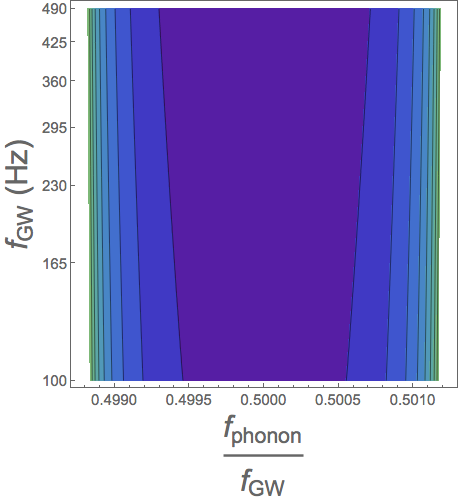}
  \caption{$c_s=0.5$ cm/s, $L=100$ $\mu$m}
\end{subfigure}
\begin{subfigure}{.3\textwidth}
  \includegraphics[width=\linewidth]{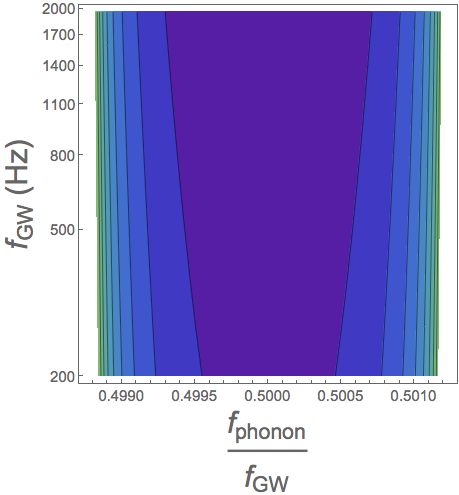}
  \caption{$c_s=1$ cm/s, $L=100$ $\mu$m}
\end{subfigure}
\begin{subfigure}{.3\textwidth}
  \includegraphics[width=\linewidth]{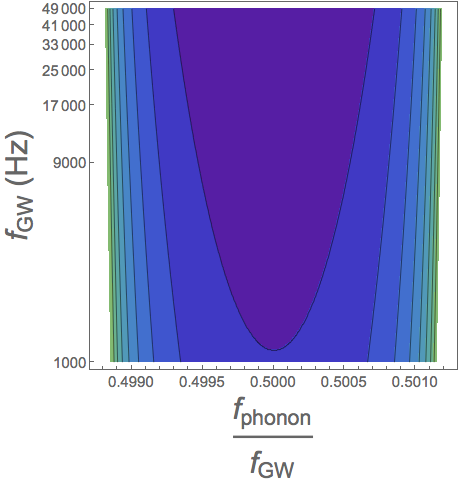}
  \caption{$c_s=5$ cm/s, $L=100$ $\mu$m}
\end{subfigure}\\
\vspace*{1cm}
\begin{subfigure}{.3\textwidth}
  \includegraphics[width=\linewidth]{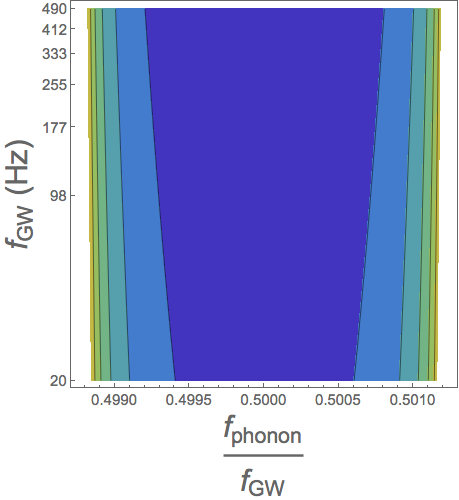}
  \caption{$c_s=0.5$ cm/s, $L=500$ $\mu$m}
\end{subfigure}
\begin{subfigure}{.3\textwidth}
  \includegraphics[width=\linewidth]{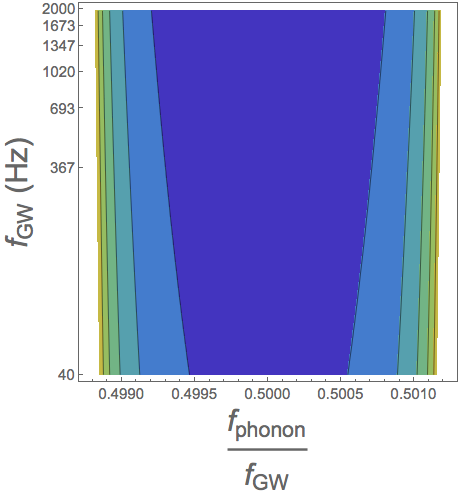}
  \caption{$c_s=1$ cm/s, $L=500$ $\mu$m}
\end{subfigure}
\begin{subfigure}{.3\textwidth}
  \includegraphics[width=\linewidth]{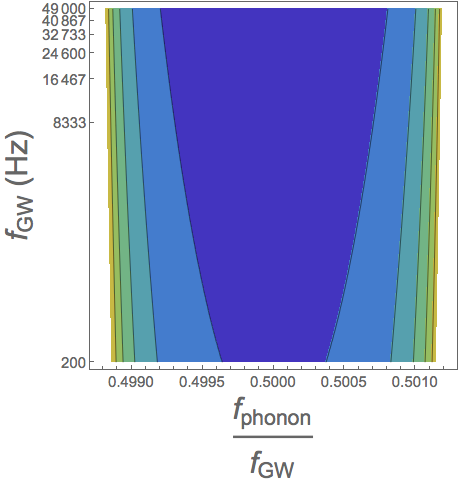}
  \caption{$c_s=5$ cm/s, $L=500$ $\mu$m}
\end{subfigure}
\caption{Dependence of the number of cycles $N$ needed to reach non-linearity, on $q=\frac{f_{\rm phonon}}{f_{\rm GW}}$ and the frequency (assuming $f_{\rm GW}=f_B$) for an undamped condensate. We see that, in general, the condensate is sensitive to a greater number of oscillations when considering a smaller frequency and larger condensates. We take the number density to be $10^{20}$ m${}^{-3}$ of ${}^{39}$K atoms. The time to non-linearity is simply $N/f_{\rm GW}$. }
\label{fig: N}
\end{figure}

\begin{figure}
    \centering
    \begin{subfigure}{0.3\textwidth}
        \includegraphics[width=\linewidth]{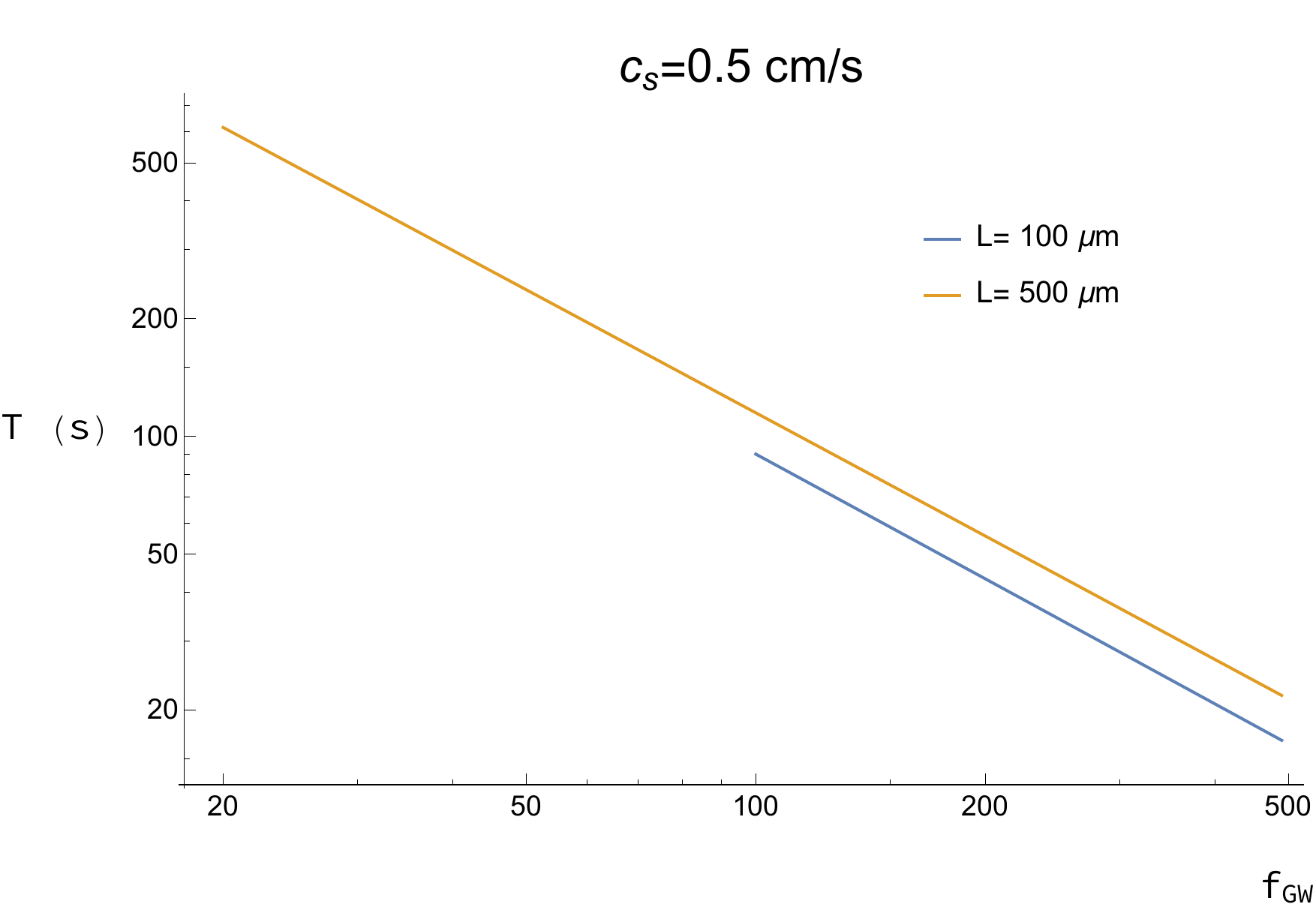}
    \end{subfigure}
        \begin{subfigure}{0.3\textwidth}
        \includegraphics[width=\linewidth]{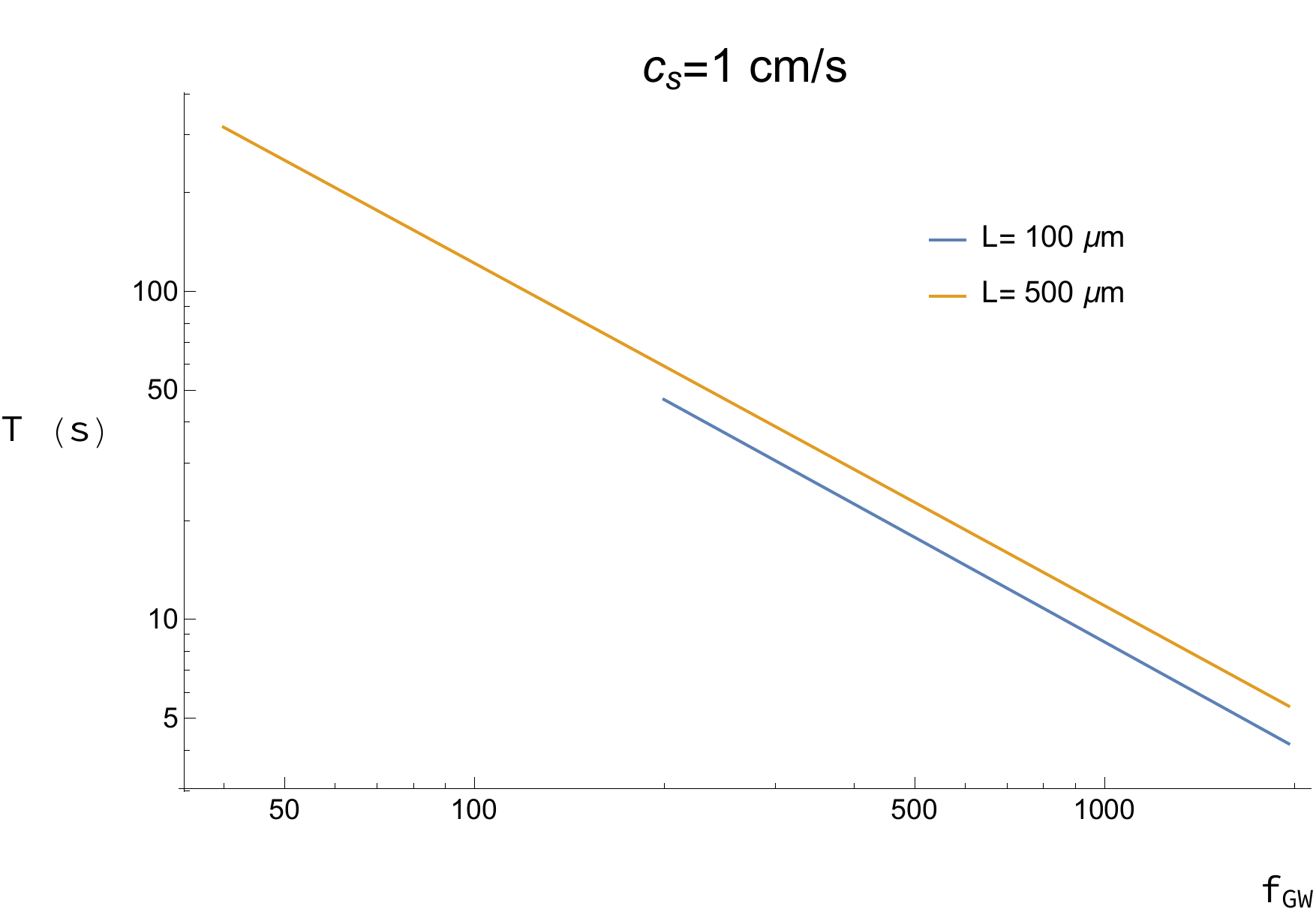}
    \end{subfigure}
        \begin{subfigure}{0.3\textwidth}
        \includegraphics[width=\linewidth]{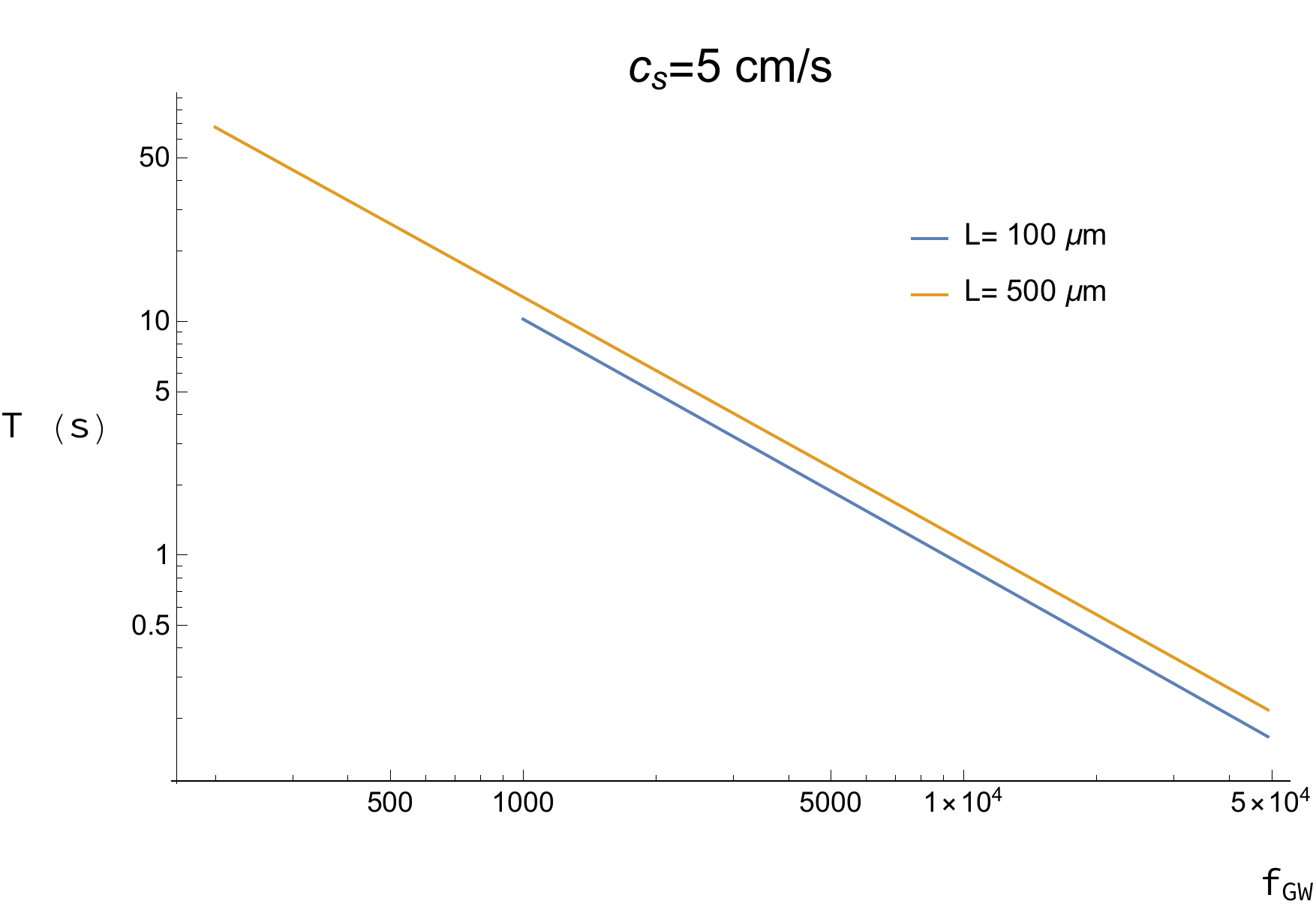}
    \end{subfigure}
    \caption{Observation time for a gravitational wave interacting with a BEC with $q=0.501245$. Lower frequencies correspond to longer observational times. We note that this neglects the effects of 3-body recombination, which limits the condensate lifetime. We discuss this effect further in Section \ref{sec: experimental}). For smaller values of $q$ (i.e. closer to the resonance peak), the maximum observational time of the condensate decreases.}
    \label{fig: Observational Time}
\end{figure}

In Figure \ref{fig: N}, we demonstrate the dependence of $N$ on $q$, the size of the condensate, and the frequency of the magnetic field. We consider a cross-section of this plot in Figure \ref{fig: Observational Time}. Throughout the remainder of this paper, all calculations and discussions will assume that this non-linearity condition is saturated.

\section{Estimating the sensitivity to gravitational waves}
\label{sec: Estimation}

As phonons are affected by the gravitational wave, the gravitational wave's amplitude is imprinted on the phonon's density matrix. The question then becomes how to extract the gravitational wave amplitude from the density matrix. This can be done by exploiting techniques in quantum metrology, which allows measurements to be done on quantities in quantum systems that are not operator observables \cite{Giovannetti2006,Nature2012,Giovannetti2012,Sabin2014b}. In this Section, we will be determine the sensitivity of an undamped condensate at zero-temperature with no extra sources of noise in the system. In Section \ref{sec: damping}, we will consider the more realistic case by investigating the sensitivity of a damped BEC. The influence of noise arising in an experimental setup will be investigated in Section \ref{sec: experimental}.

Given a parameter $\tilde\epsilon$ in a quantum system, the quantum Cramer-Rao bound is \cite{Braustein1994}
\begin{align}
\braket{(\Delta \epsilon)^2}\geq\frac{1}{MH_\epsilon}\ ,
\end{align}
where $\braket{(\Delta \epsilon)^2}$ is the expectation value of the uncertainty in $\epsilon$, $M$ is the number of measurements of the system, and $H_\epsilon$ is the Fisher information  \cite{Braustein1994}
\begin{align}
H_{\epsilon}=\frac{8\bigg(1-\sqrt{F(\rho_{\epsilon},\rho_{\epsilon+d\epsilon})}\bigg)}{d\epsilon^2}\ ,
\label{eq: QFI}
\end{align}
with
 $F(\rho_{\epsilon},\rho_{\epsilon+d\epsilon})=\left[\Tr\sqrt{\rho_\epsilon\sqrt{\rho_{\epsilon+d\epsilon}}\rho_\epsilon}\right]^2$ quantifying the overlap (fidelity) between the states $\rho_{\epsilon}$ and $\rho_{\epsilon+d\epsilon}$ \cite{Uhlmann1976,Jozsa1994}. We find the sensitivity to gravitational waves is 
 
 \begin{align}
 \braket{(\Delta\tilde\epsilon)^2}=\frac{4}{M\left(\Re\left[\alpha_{rel}^{(1)}\right]+|\beta_{rel}|\right)^2\left(\cosh^4r_0+\sinh^4r_0\right)}\ .
 \label{eq: sensitivity}
 \end{align}
 where $r_0$ is the initial squeezing of the phonons, to $\alpha_{rel}$ in the gravitational wave amplitude, with $\alpha_{rel}$ and $\beta_{rel}$ given by  equations \eqref{eq: alpha rel} and \eqref{eq: beta rel}, respectively. The details of this calculation are provided in Appendix \ref{sec: derivation}.
 
In this and all subsequent considerations, we take $M=1$. This corresponds to a gravitational wave interacting with a $M$ BECs for $N$ periods. We also assume that the BEC can continuously be regenerated over the course of a year\cite{Streed2006,Tiecke2009}, as suggested by \cite{Sabin2014}, such that observations can be continuously made of a gravitational wave, whose frequency remains approximately constant over the total time of observation. In this case, we find

\begin{align}
 \sqrt{\braket{(\Delta\tilde\epsilon)^2}}=\sqrt{\frac{1}{N_{tot}}}\frac{2}{\left(\Re\left[\alpha_{rel}^{(1)}\right]+|\beta_{rel}|\right)\sqrt{\cosh^4r_0+\sinh^4r_0}}\ ,
 \label{eq: sensitivityYear}
 \end{align}
 where $N_{tot}$ is the total number of regenerations of the condensates. For simplicity, we consider a year of continuous observation, so $N_{tot}\approx\frac{\rm{1 yr}}{N/f}$.
 
By expanding \eqref{eq: sensitivityYear} around $q=\frac{1}{2}$, and taking $N$ to be large, we find that

\begin{align}
\sqrt{\braket{(\Delta\tilde\epsilon)^2}}\sim \frac{384}{\sqrt{N_{tot}}} a^2 e^{a \left(\frac{5 \pi }{2}-\pi  N\right)}\ ,
 \end{align}

We note that, in practice, it would be quite difficult to reconstruct the BEC with exactly the same properties. Extra source noise would be introduced into the system, and so such a BEC machine would be beneficial only if the noise resulting from an inexact replication is smaller than the gain in sensitivity resulting from multiple BEC experiments.

We also note that, in equation \eqref{eq: sensitivityYear}, we are calculating the sensitivity of the scaled gravitational wave amplitude $\tilde\epsilon$, rather than $\epsilon$ itself. To obtain $ \braket{(\Delta\epsilon)^2}$ from  $\braket{(\Delta\tilde\epsilon)^2}$, one can either average or maximize over the components $k_{x,y}$ of the wavevector. As both methods scale the estimation of the gravitational wave amplitude by $\mathcal{O}(1)$, we will not include this effect in our subsequent calculations and Figures.

\begin{figure}
    \centering
    \includegraphics[scale=0.4]{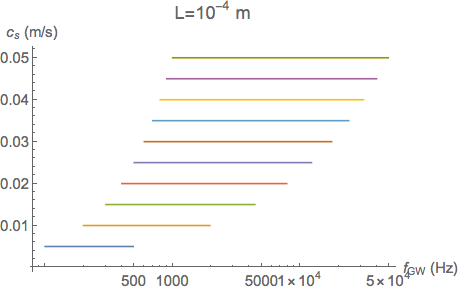}
    \includegraphics[scale=0.4]{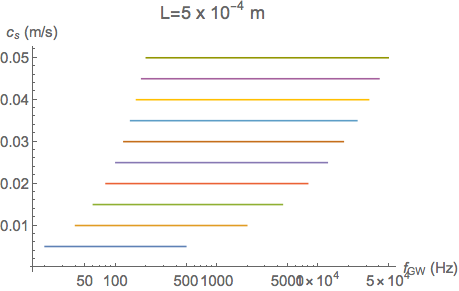}
    \caption{Frequency regime of a BEC gravitational wave detector for ${}^{39}$K atoms. Note that the larger condensates can probe lower-frequency gravitational waves, whereas those with faster speeds of sound can probe higher-frequency gravitational waves.}
    \label{fig: FrequencyRegime}
\end{figure}

As the BEC has a finite size, we note that the minimum gravitational wave frequency that can be observed is $f\approx\frac{2c_s}{L}$ (where we use the relation that resonance between the gravitational wave and phonon occurs at $q=\frac{\omega}{\Omega}\approx\frac{1}{2}$ and $c_s$ is the speed of the phonons). The maximum frequency observed is derived from the chemical potential $\mu=mc_s^2$, which sets the upper bound on the frequency of phonons created: $f\ll\mu$. Again using $q\approx\frac{1}{2}$, we see that the maximum gravitational wave frequency is $f\approx\frac{2mc_s^2}{2\pi\hbar\times10}$, where the $10$ comes from the inequality. We show in Figure \ref{fig: FrequencyRegime} how the frequency bands of our gravitational wave detector are dependent on the speed of sound and length of the condensate. For lower-frequency gravitational wave detection, a larger condensate is required while high-frequency gravitational wave detection requires a faster speed of sound.

\begin{figure}
\centering
\begin{subfigure}{.45\textwidth}
  \centering
  \includegraphics[width=\linewidth]{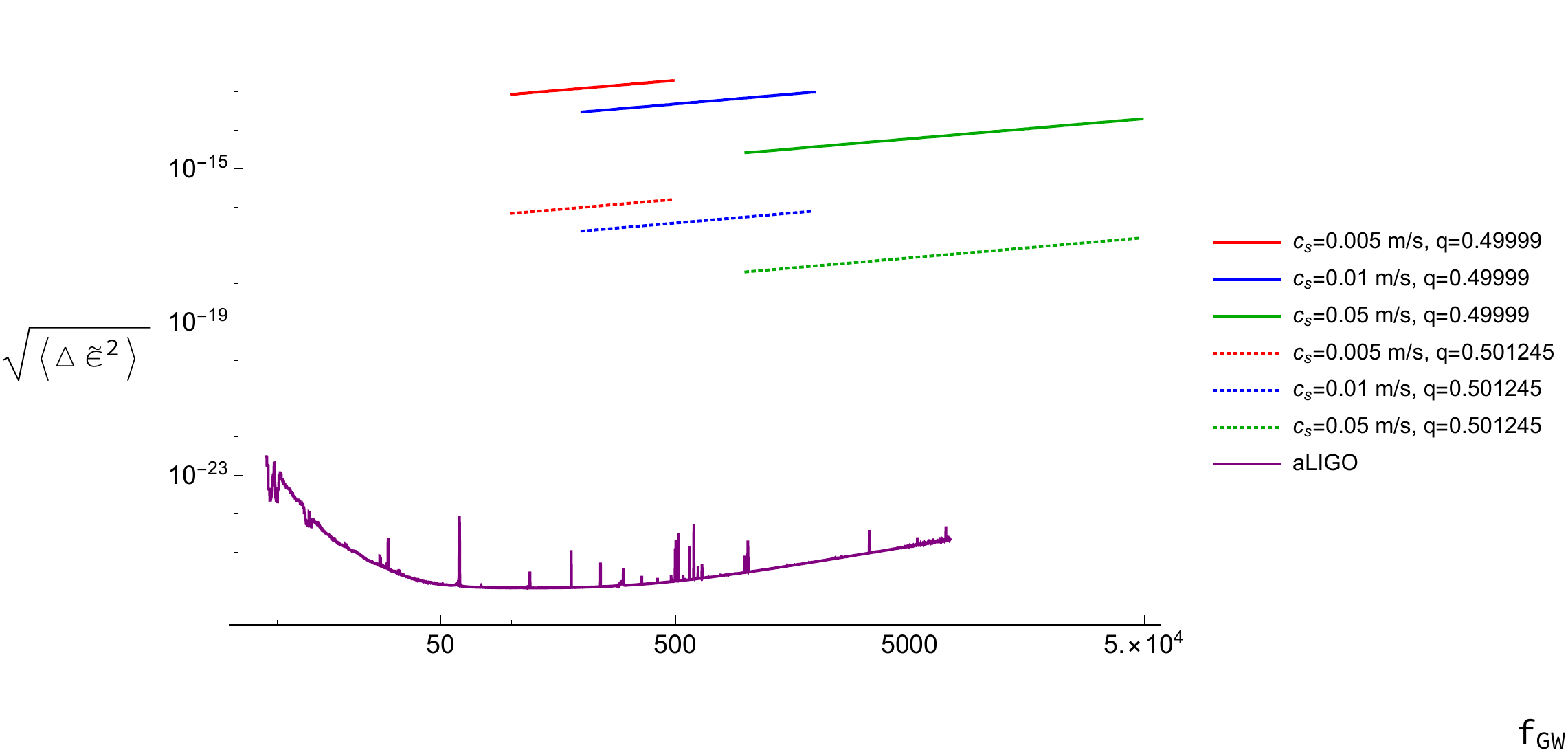}
  \caption{$L=100$ $\mu$m}
\end{subfigure}
\begin{subfigure}{.45\textwidth}
  \centering
  \includegraphics[width=\linewidth]{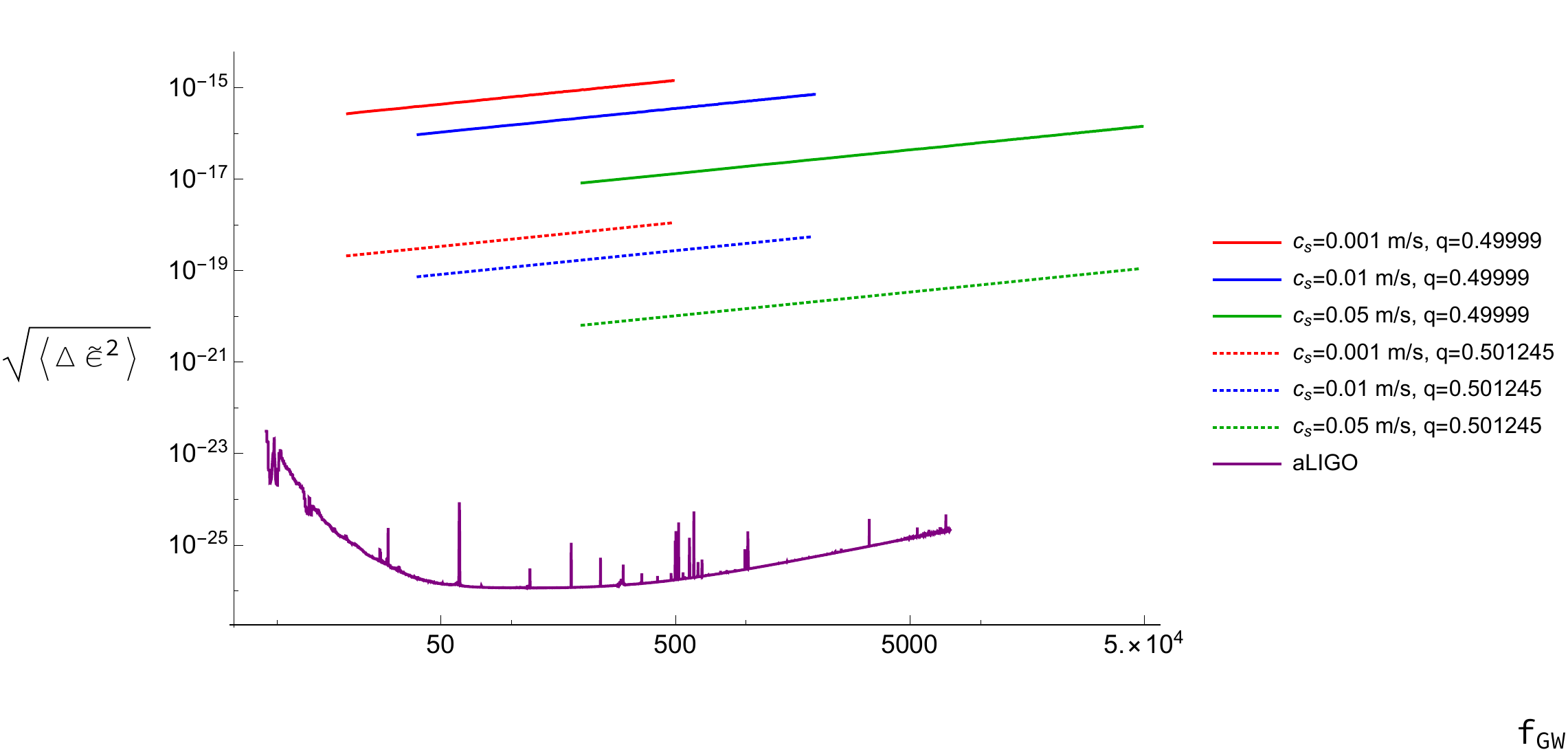}
    \caption{$L=500$ $\mu$m}
\end{subfigure}
  \caption{The strain sensitivity of an undamped BEC gravitational wave when on resonance with an oscillating magnetic field. In each case, the condensate is observed for the maximum number of oscillations $N(q)$. We conduct the hypothetical experiment over the course of a year. We take the number density to be $10^{20}$ m${}^{-3}$ of ${}^{39}$K atoms).}
  \label{fig: sensitivity undamped}
\end{figure}

\begin{figure}[t!]
    \begin{subfigure}{\textwidth}
\centering
\includegraphics[scale=0.7]{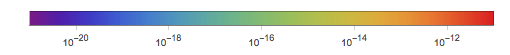}
\caption*{BEC sensitivity to gravitational waves.}
\end{subfigure}\vspace{0.5cm}
\begin{subfigure}{.3\textwidth}
  \includegraphics[width=\linewidth]{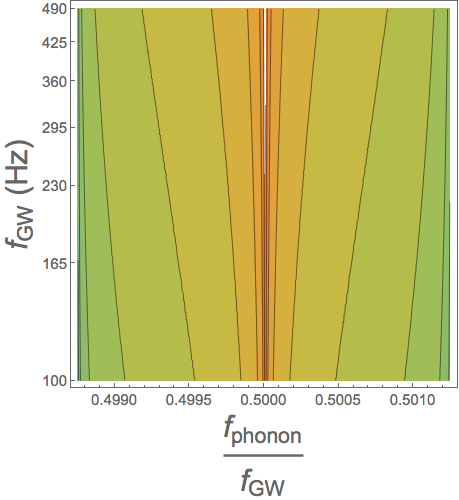}
  \caption{$c_s=0.5$ cm/s, $L=100$ $\mu$m}
\end{subfigure}
\begin{subfigure}{.3\textwidth}
  \includegraphics[width=\linewidth]{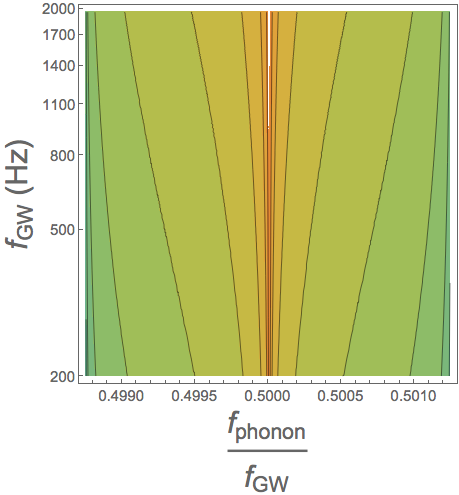}
  \caption{$c_s=1$ cm/s, $L=100$ $\mu$m}
\end{subfigure}
\begin{subfigure}{.3\textwidth}
  \includegraphics[width=\linewidth]{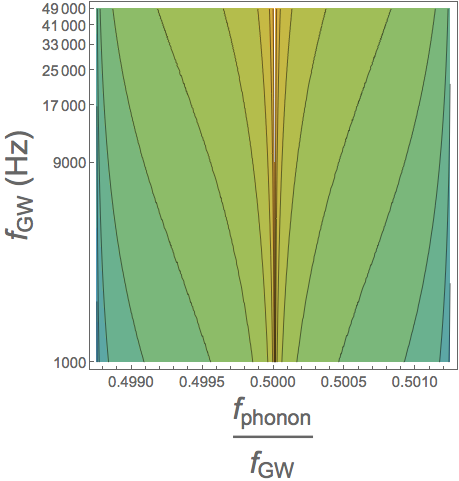}
  \caption{$c_s=5$ cm/s, $L=100$ $\mu$m}
\end{subfigure}\\
\vspace*{1cm}
\begin{subfigure}{.3\textwidth}
  \includegraphics[width=\linewidth]{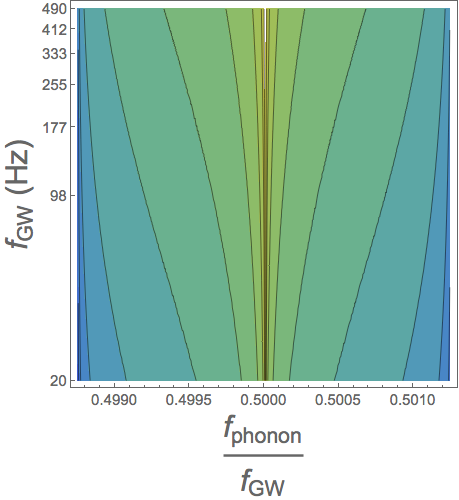}
  \caption{$c_s=0.5$ cm/s, $L=500$ $\mu$m}
\end{subfigure}
\begin{subfigure}{.3\textwidth}
  \includegraphics[width=\linewidth]{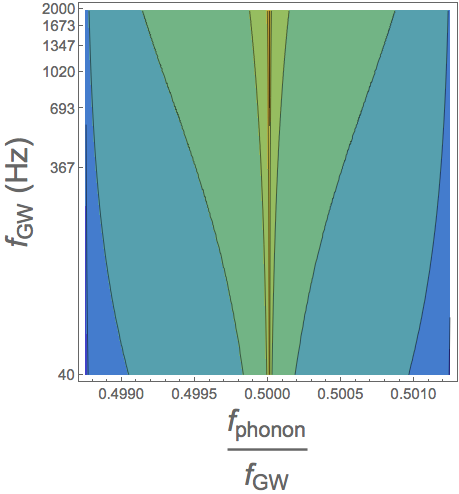}
  \caption{$c_s=1$ cm/s, $L=500$ $\mu$m}
\end{subfigure}
\begin{subfigure}{.3\textwidth}
  \includegraphics[width=\linewidth]{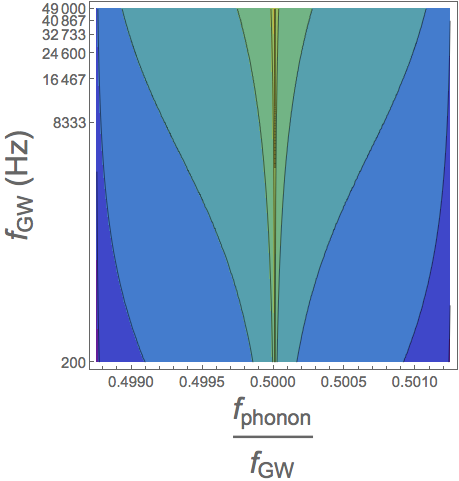}
  \caption{$c_s=5$ cm/s, $L=500$ $\mu$m}
\end{subfigure}
\caption{Contour plot of the strain sensitivity of an undamped BEC to gravitational waves. The condensate is observed for the maximum number of oscillations $N(q)$. We run the experiment over the course of a year and see that undamped condensate could be sensitive to gravitational waves across the frequency spectrum. Changing the speed of sound and length of the condensate will affect the minimum and maximum gravitational wave 
frequency
that can be observed. We take the number density to be $10^{20}$ m${}^{-3}$ of ${}^{39}$K atoms).}
\label{fig: sensitivity undamped contour}
\end{figure}

To illustrate the sensitivity to gravitational waves, we take the number density to be $10^{20}$ m${}^{-3}$ and a mass of the atoms to be $10^{-25}$ kg. In Figures \ref{fig: sensitivity undamped} and  \ref{fig: sensitivity undamped contour}, we consider the sensitivity of an (undamped) BEC and illustrate the effect of $q$, length of the condensate, and frequency of incoming gravitational waves on strain sensitivity. We note that a BEC in the presence of an oscillating magnetic field is most sensitive to higher frequency gravitational waves, similar to \cite{Sabin2014} and \cite{Robbins2019}. In Section \ref{sec: damping}, we consider a more realistic situation by modelling Beliaev damping of the phonons within the condensate \cite{Griffin2009}.

This effect of having greater sensitivity away from resonance must be highlighted. When the system is away from $\frac{\omega}{\Omega}=\frac{1}{2}\pm\delta$, where $\delta\ll1$, we saw that the sensitivity to gravitational waves is actually increased. In these off-resonance cases, it takes longer for the system to become non-linear (as the phonons have less energy), hence we can observe gravitational waves for a longer time (a cross-section of this effect was illustrated in Figure \ref{fig: sensitivity undamped}).

\section{Damping}
\label{sec: damping}
In Section \ref{sec: Estimation}, we considered an undamped BEC with no additional sources of noise and demonstrated that, if such a system could be created, then a BEC could potentially be used to detect gravitational waves across several orders of magnitude in frequency, depending on the speed of the phonons and the length of the condensate. In reality, even at zero-temperature (which was implicitly considered), the phonons within the BEC will naturally undergo decoherence through Beliaev damping, where the damping rate is given by  \cite{Giorgini1998}
\begin{align}
\gamma_B=\frac{3}{640\pi}\frac{\hbar\omega^5}{mnc_s^5}\,
\end{align}
for a cubic BEC with no magnetic field,
where any magnetic field correction would manifest itself as a higher-order term because $\frac{\delta B}{B}\ll1$. 

Since the density matrix is a Gaussian state, we can use a covariance matrix to compute the Fidelity. As demonstrated in \cite{Howl2018}, the covariance matrix $\sigma$ is damped as $\sigma= e^{-\gamma_Bt}\sigma_0+(1-e^{-\gamma_Bt})\sigma_{\infty}$, where $\sigma_0$ is the initial covariance matrix and $\sigma_{\infty}$ is the $t\to\infty$ covariance matrix. For our purposes, we can neglect the second term as $\gamma_B t\gg1$. From \cite{Ahmadi2014}, the quantum Fisher information depends only the square of several combinations of elements of the covariance matrix. Hence, the quantum Fisher information is damped as $H_{\epsilon,\text{damped}}\sim e^{-2\gamma_Bt}H_{\epsilon,\text{undamped}}$. Therefore,
\begin{align}
 \braket{(\Delta\tilde\epsilon)^2}=\frac{1}{T_{tot}}\frac{4}{M\left(\Re\left[\alpha_{rel}^{(1)}\right]+|\beta_{rel}|\right)^2\left(\cosh^4r_0+\sinh^4r_0\right)}e^{2\gamma_Bt}
\end{align}
where $t=N_{max}T$ is the maximum time of observation of the condensate, $N$ is the maximum number of oscillations of the magnetic field, and $T=\frac{2\pi}{\Omega}$ is the period of the gravitational wave. We note that $N_{max}T<\frac{1}{\gamma_B}$, where $\frac{1}{\gamma_B}$ is approximately the decoherence time \cite{Howl2018}.

\begin{figure}
\centering
\includegraphics[scale=0.4]{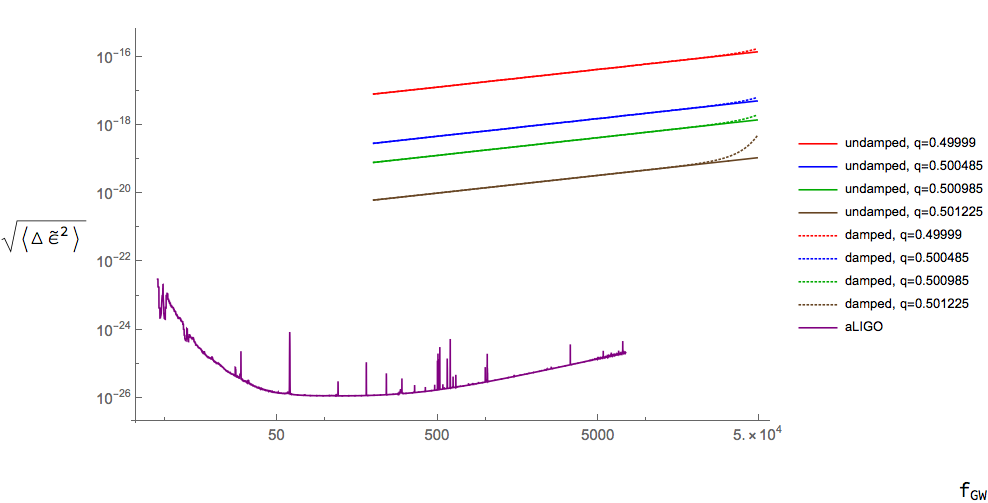}
  \caption{The strain sensitivity of a damped BEC gravitational wave detector for various values of $q$ (=ratio of phonon to GW frequency). We consider $c_s=5$ cm/s and $L=500$ $\mu$m. In each case, the condensate is observed for the maximum number of oscillations $N(q)$, before non-linearities become important. We run the experiment over the course of a year and see that damping as a negligible effect for a potassium-39 BEC whose number density is $10^{20}$ m${}^{-3}$.}
  \label{fig: qfLdamped}
\end{figure}

In Figure \ref{fig: qfLdamped}, we consider how the strain sensitivity is affected when damping is considered for various gravitational wave frequencies and values of $q$. We see that larger condensates and faster speeds of the phonons give rise to greater sensitivity. However, we note that the sensitivity of a BEC to gravitational waves is optimal at lower frequencies, in contrast to the results of \cite{Sabin2014} and \cite{Robbins2019}. In Figure \ref{fig: qfLdamped}, when fixing $q$, we see that the the strain sensitivity becomes worse further away from the resonance peak of $q=\frac{1}{2}$. However, for a ${}^{39}K$ BEC, the effect of damping is negligible for most values of $q$.

\subsection{Squeezing}

 In our previous work \cite{Robbins2019}, we noted that squeezing the phonon states seemed to be a necessary feature for a BEC to detect gravitational waves. A similar conclusion was implied by the figures in \cite{Sabin2014}: to exceed the sensitivity of LIGO, the phonons needed to be squeezed by much more than the capabilities of current experimental designs (7.2 dB), though it was noted that there might be a means \cite{Serafini2009} of exceeding current limitations.
 
Though squeezing improves the sensitivity to gravitational waves in our current work by a factor of 
$$\sqrt{\sinh^4r_0+\cosh^4r_0},
$$
we note that this feature is no longer an essential element of our proposal. For $r=0.83$ (potentially achievable with current technology \cite{Gu2014}), this improves sensitivity by only a factor of $2.05$. Improvements in squeezing may increase the sensitivity to gravitational waves, though we note that squeezed states would contribute to the energy density of the condensate, and therefore cause the condensate to become unstable quicker (Section \ref{sec: Non-linearities}. Furthermore, the  squeezing parameter would also decay, as discussed in \cite{Howl2018}.

At the present time using current techniques (which we discuss in more detail in the next section), it is not feasible to detect gravitational waves using a BEC, though the results of this paper indicate that this could potentially be achieved in the future. When this occurs, then Beliaev damping and the effects of squeezing may need to be considered in more detail. However, in the case of observing a continuous gravitational wave by using parametric resonance, squeezing and Beliaev damping is not the primary limiting factor as  suggested in \cite{Robbins2019}. Rather it is 3-body recombination, which we discuss in the following section.

\section{Experimental Implementation}
\label{sec: experimental}

\subsection{Observable Sources}
\label{sec: observable}

In Figure \ref{fig: qfLdamped}, we demonstrated that with a speed of sound of $c_s=5$ cm/s and a cubic condensate with side lengths of $L=500$ $\mu$m, gravitational waves of amplitude $\mathcal{O}(10^{-20})$ could be detected at frequencies around 500 Hz. In Figure \ref{fig: sensitivity undamped}, lower frequency gravitational waves could be detected if their amplitude was $\mathcal{O}(10^{-21})$ and higher. However, as was noted in \cite{Riles2017}, the amplitude of continuous gravitational waves tends to be weaker than that of transient gravitational waves.

\begin{figure}[h!]
    \centering
    \includegraphics{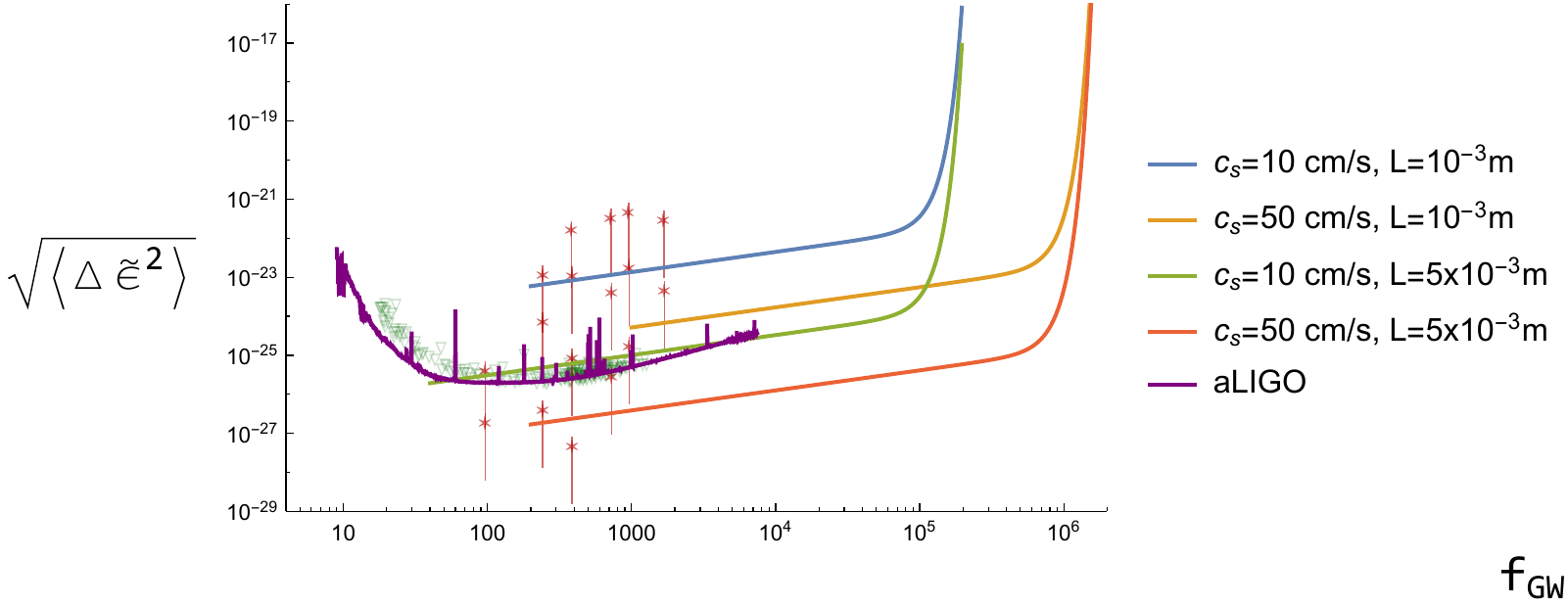}
    \caption{Speeds of sound and length of the condensate necessary to observe gravitational waves with a smaller amplitude. The green triangles represent the upper bounds the strain  of various pulsar sources \cite{Abbot2017} while the red stars correspond to axion clouds based on several scenarios proposed in \cite{Zhu2020}. The error bars for the axion clouds assume a distance of 1kPc - 100 kPc.}
    \label{fig: Advanced}
\end{figure}

For this reason, it is unlikely that a BEC could detect gravitational waves without going to larger sizes or faster speeds of sound. In Figure \ref{fig: Advanced}, we illustrate the speeds of sound and condensate lengths necessary to detect gravitational waves with a smaller amplitude, where we also account for Beliaev damping. Currently, it is not possible to create condensates that large, nor is it possible for the condensates to have a long enough lifetime; as we discuss in the next section, larger speeds of sound increase the three-body recombination rate, which in turn decreases the lifetime of the condensate.

A Bose-Einstein condensate is an advantageous gravitational wave detector as it is capable, in principle, of observing gravitational wave sources across several orders of magnitude in frequency, depending on the condensate's length and speed of sound. At gravitational wave frequencies in the tens or hundreds of Hertz, continuous signals from rotating neutron stars or pulsars could be detected \cite{Abbott2017,Covas2020}. The amplitude of such waves may be $\mathcal{O}(10^{-24})$. For newly-created fast-spinning magnetars, the frequency of gravitational waves could be 0.5-2 kHz, with amplitudes as large as $\mathcal{O}(10^{-21})$, though we note that such signals will only be continuous on the order of weeks, rather than years. \cite{Stella2005}

Around black holes, it has been theorized that a boson cloud could form through superradiance, and then emit continuous gravitational waves during annhilation processes. \cite{Zhu2020}
The frequency of the emitted waves will be dependent on the mass of the bosons and distance from Earth, though could potentially be in the kilohertz regime (smaller boson masses will have frequencies in the tens or hundreds of Hertz). The amplitudes of such waves could be as large as $\mathcal{O}(10^{-22})$ \cite{Zhu2020}. One candidate for boson clouds is axions. In addition to annihilations, axions can undergo energy level transitions, which will also emit gravitational waves. These tend to be in the low-frequency regime ($<200$ Hz) and have amplitudes $\mathcal{O}(10^{-24})$ and below, though one energy level transitions could be $\mathcal{O}(10^{-22})$ (frequency of gravitational wave $\sim 40$ Hz). At higher frequencies, there may be additional cosmological sources, such as cosmic strings or inflationary signatures. However, we note that the signals from such sources may be too weak to be detected by our proposed setup \cite{Battye1998}.

In Figure \ref{fig: Advanced}, we also illustrate the amplitude and frequency regimes of such signals. We find that, though it may be necessary to go to larger length scales and speeds of sound to observe most of the continuous gravitational wave sources, it may still be possible to detect gravitational waves using condensates of lengths $L\sim 5\times10^{-4}$ m and speeds of sound $c_s\sim5\times10^{-2}$ m/s. 

\subsection{Experimental Design}

We have presented the theoretical sensitivity for a BEC gravitational wave detector in the ideal case and in the more realistic case of zero-temperature (Beliaev) damping. We saw that Beliaev damping will decrease sensitivity to gravitational waves at higher frequencies and smaller speeds of sound. At lower frequencies, Beliaev damping is no longer relevant (though Landau damping could become important at non-zero temperatures) \cite{Giorgini1998}. As we will discuss in this section, with current or reasonably projected technologies, detection will be limited by the noise in the detector (though the precise source of the noises will be dependent on the experimental setup).

Bose-Einstein condensates have been previously confined to cylindrical boxes using optical traps \cite{Meyrath2005,Gaunt2013}. For simplicity, our theoretical analysis in the preceding sections considered a cubic box trap but the exact geometry is not critical to the results. A cylindrical trap can be engineered to have a phonon resonance at any frequency available to a cubic box. In \cite{Gaunt2013}, such a potential was generated with a radius of $15\pm1$ $\mu$m and length of $63\pm2$ $\mu$m. The ``flatness'' of the box corresponded to a potential $\propto r^{13\pm 2}$. We envision a similar structure of our potential. 

The simplest way to modulate the length of the trap is to modulate the power of the trapping laser. This leverages the fact that the potential has the shape of a high-order polynomial rather than a perfectly sharp edge. Such a technique has been used to excite sound waves in Fermi gases in a box trap \cite{Zwierlein2019}.

We envision the trap to be created by suspending the optics for generating the trap beams using a similar method as used for suspending the large mirrors in LIGO \cite{Martynov2016}. By assuming that the noise within the BEC gravitational wave detector follows a similar spectrum as that of LIGO, we can use LIGO's sensitivity curves to estimate the impact of vibrational noise on the BEC. We also note that, similar to \cite{Gaunt2013}, an additional, relatively small magnetic field gradient will need to be applied to cancel out the Earth's gravity. 

Before discussing the effects of vibrations, we discuss the dominant source of noise: laser intensity noise. The length of a box trap, in the Thomas-Fermi approximation, is the distance between the points at the two ends where the trapping potential equals the chemical potential. This can be modulated by changing the positions of the light sheets that form the hard walls at either end of the box or by modulating the intensity of the trapping light. To analyze the impact of light intensity noise,  we consider the equation $\mu = U_0 e^{-2(z/w_0)^2}$, where $\mu$ is the chemical potential, $U_0$ is the peak potential for a blue-detuned light sheet, $w_0$ is the Gaussian beam waist, and $z$ is the distance from the intensity peak to the edge of the trap defined by the equation. Since $U_0$ is proportional to the power $P$ in the light sheet, we note that $dz/d{U_0} = w_0^2/(4zU_0)$ implies the change in z, $\delta z$, due to a small change in power, $\delta P$, is 
\begin{equation}
	\delta z = \frac{w_0^2}{4 z} \frac{\delta P}{P}.
\end{equation} 
For reasonable laser powers and focal parameters, $z\approx 2w_0$. For a cubic box of length $L$, the minimum waist is roughly found by setting the Rayleigh range equal to the length: $L = \pi w_0^2/\lambda$, which gives $\delta z = \sqrt{L \lambda}/(8 \sqrt{\pi}) \cdot \delta P/P$, Assuming correlated power fluctuations for the sheets at both ends of the trap (as from laser power noise; fluctuations due to beam splitting to make the two sheets should be anti-correlated and so cancel to leading order) we find finally
\begin{equation}
	\frac{\delta L}{L} = \sqrt{\frac{\lambda}{16 \pi L}} \frac{\delta P}{P}.
\end{equation} 

Using $532\;{\rm nm}$ light as in \cite{Gaunt2013} and a length of $1\; {\rm mm}$ gives a sensitivity $\delta L/L = 0.003  (\delta P/P)$. The current state of the art for this wavelength could allow for $50\;{\rm W}$ beams. In an experiment lasting three seconds, assuming the power is controlled to the shot noise limit, we find the length noise is $\delta L/L = 1.5\times 10^{-13}$. This sets a bound on the strain sensitivity for such a technique in a single experiment. 

\begin{figure}[t!]
    \centering
    \includegraphics[scale=0.7]{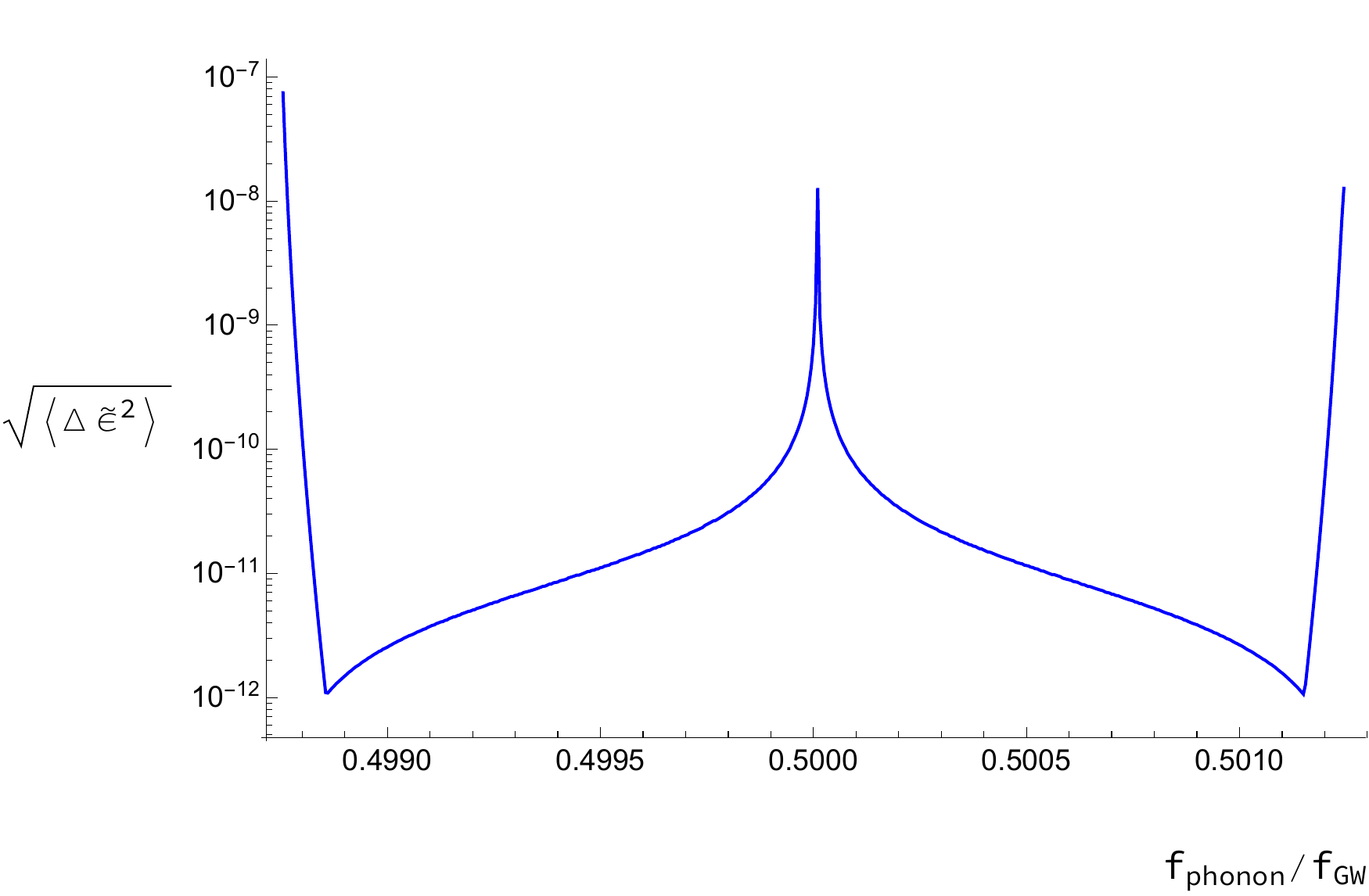}
    \caption{The sensitivity of a BEC ($a=0.005$) to gravitational waves, assuming that the BEC has side lengths of $100$ $\mu$m, the speed of sound is $1$ cm/s, the lifetime of the BEC is limited to 3 s (without using a BEC machine), and the incoming gravitational wave has a frequency of $1000$ Hz. We see that the greatest sensitivity to gravitational waves is achieved slightly off the centre of resonance and that moving outside the resonance band destroys the sensitivity to gravitational waves.}
    \label{fig: realistic sensitivity}
\end{figure}

In Figure \ref{fig: realistic sensitivity}, we depict the sensitivity curve of a BEC to a 1000 Hz gravitational wave, where the observation time is taken to be the minimum of the non-linearity time and the lifetime of the BEC (3 seconds). The minimum of the sensitivity curve corresponds to the location where the non-linearity time is exactly equal to the lifetime of the BEC. Moving outside the resonance bands causes the non-linearity time to exceed the lifetime of the condensate, which implies that there is insufficient time to use parametric resonance to amplify the sensitivity. In this case, such a BEC would be insensitive to gravitational waves.

This intrinsic BEC noise can potentially be improved by squeezing the intensity noise of the laser, since the phase of the trapping light is not critical. State of the art squeezing achieves $15\;{\rm dB}$ reduction in noise \cite{Squeeze1,Squeeze2}. Supposing a substantial improvement to $30\;{\rm dB}$ is possible in the future, the achievable strain sensitivity of $\sim 10^{-16}$ is still far behind the sensitivity of LIGO. Using an asymmetric box potential and only looking at phonon modes along the long direction would allow for larger $L$ and a smaller beam waist for the walls in the important direction. A $5\;{\rm mm}$ long trap with transverse width of $100\;{\rm \mu m}$ could improve the sensitivity by nearly an order of magnitude. We conclude that the limit imposed by plausible near-term technologies is a strain sensitivity of $10^{-17}$ per shot.  

Mechanical vibrations of the optical setup are the other potential source of trap length noise. To minimize the impact of vibrations, we assume the experiment is performed in a large ultrahigh vacuum chamber with optics built directly into a single piece of Zerodur glass \cite{Sengstock2014} suspended on a vibration control system with comparable specifications as the LIGO mirror suspension system \cite{Martynov2016}. A phonon mode excited by gravitational waves would have different symmetry than the phonon modes that may be excited by center of mass oscillations of the trap. However, the frequencies of interest will dominantly transfer as center of mass oscillations. The strong mismatch between internal vibrational excitation resonances and the pendular modes of the optical system as a whole mean the transfer function from vibrations transmitted through the suspension system to length fluctuations will be much less than one. Assuming the transfer function is 1, we find a strain sensitivity of $10^{-17}$. Given that the transfer function will be well below this level, we find that even with current technology vibrational noise can be suppressed below the limits imposed by laser intensity fluctuations. 

These limits are general to any technique that would use the modulation of a BEC trap due to a gravitational wave as a signal. The uncontrolled modulations of trap length due to laser power fluctuations or vibrations provide a noise foreground that must be overcome for any of the proposed schemes to be experimentally viable \cite{Sabin2014,Robbins2019}.

The ultimate objective is to measure the quantum Fisher information. This can be calculated by representing the quantum Fisher information in terms of the covariance matrix. For a Gaussian state, the covariance matrix is $\sigma_{mn}=\frac{1}{2}\braket{X_mX_n+X_nX_m}-\braket{X_m}\braket{X_n}$, where $X_{2n-1}=\frac{1}{\sqrt{2}}(a_n+a_n^{\dagger})$, $X_{2n}=\frac{1}{\sqrt{2}i}(a_n-a_n^{\dagger})$, and $a_n,a_n^{\dagger}$ are the creation and annihilation operators. Let us also assume $\braket{X_i}=0$. Let $\sigma_A$ be the Gaussian state of the phonons before interaction with the gravitational wave and $\sigma_B$ be the Gaussian state of the phonons afterwards. Define  \cite{Marian2012}
\begin{subequations}
\begin{align}
\Delta&=\det[\mathcal{\sigma}_A+\mathcal{\sigma}_B]\ ,\label{eq: Delta}\\
\Lambda&=2^{2n}\det\left[\mathcal{\sigma}_A+\frac{i}{2}J\right]\det\left[\mathcal{\sigma}_B+\frac{i}{2}J\right]\ ,\label{eq: Lambda}\\
J&=\bigoplus_{k=1}^n\begin{pmatrix}
0 & 1\\
-1 & 0
\end{pmatrix}\ ,
\end{align}
\label{eq: DLJ}
\end{subequations}
where $n$ is the number of modes. The fidelity of the two single-mode states is then given by \cite{Marian2012}
\begin{align}
F(\sigma_A,\sigma_B)=\frac{1}{\sqrt{\Delta+\Lambda}-\sqrt{\Lambda}}\ .
\label{eq: F}
\end{align}
and using equation \eqref{eq: QFI} yields the quantum Fisher information. Thus, measuring the quantum Fisher information reduces to a determination of the individual elements of the covariance matrix. However, doing such measurements will only be possible when the intrinsic noise of the BEC is sufficiently reduced. 

Even if the technical noise of the trap is mitigated, there will remain a few caveats. In Section \ref{sec: damping}, we noted that faster speeds of sound gave rise to greater sensitivity. This was because the damping rate was $\gamma_B\propto c_s^{-5}$. With faster speeds of sounds, a BEC will have a larger minimum observable gravitational wave frequency ($\propto c_s$), though a much larger maximum observable gravitational wave frequency ($\propto c_s^2$). However, at larger speeds of sound, we note that the maximum time of observation will no longer be given by the non-linearity condition of Section \ref{sec: Non-linearities}, but by three-body recombination. When three atoms collide, two are able to form a bound, molecular state with the third escaping the system at a rate given by $\dot{n}\propto n^3(t)$, where $n(t)$ is the number density and the proportionality constant describes the rate of recombination and depends upon the properties of the system \cite{Zaccanti2009,Sorensen2013}. Larger speeds of sound correspond to greater number densities, implying a greater decrease in the number atoms in the condensate \cite{Lee1959,Meppelink2009}. We also note quantum depletion, which describes the loss of atoms from the condensate because of interactions between the atoms and excitations, will contribute to additional noise in the condensate, where the number of atoms in an excited state (therefore, leaving the condensate) is given by \cite{Pethick2008}
	\begin{align}
		\frac{n_{ex}}{N}=\frac{8}{3\sqrt{\pi}}(na^3)^{1/2}\,
	\end{align}
where $a$ is the scattering length and is valid for a zero-temperature uniform Bose gas.

\section{Conclusion and Future Prospects}
\label{sec: Conclusion}

We have considered a Bose-Einstein condensate in a trap with oscillating length to investigate its sensitivity towards a continuous gravitational wave of the form $h_{+}=\epsilon\sin\Omega t$. We showed that the oscillation gives rise to parametric resonance in the system, which can be exploited to improve the sensitivity to gravitational waves at lower frequencies compared to higher frequencies.

Despite the attractive aspects of this idealized setup, this proposal has formidable challenges in its actual experimental implementation.
The technical noise present using current technology would overwhelm signals from likely gravitational wave sources. The finite lifetime of real BECs also imposes a difficult constraint, which we implemented as a hard limit of 3 seconds on the length of an experiment. However, if the technical or lifetime issues can be overcome, our results indicate that parametric resonance can make a BEC a more promising gravitational wave detector than earlier considerations have indicated \cite{Schutzhold2018,Robbins2019}.

We offer three prospects for improving the sensitivity beyond the limitations discussed on the paper. Three-body loss coefficients are relatively similar amongst laser coolable atoms. The one atom that has produced a BEC that could potentially have a dramatically longer lifetime is hydrogen. Since the triplet potential of ${\rm H}_2$ supports no bound states, there is no three-body loss in spin-polarized H. Hydrogen BEC experiments in magnetic traps suffer from dipolar relaxation \cite{Kleppner1998}. In principle, H polarized in the lowest-energy Zeeman sublevel could produce BECs with very long lifetimes. This state could be trapped in an optical potential, though using repulsive light sheets as described in the paper would be infeasible, given the unavailability of strong laser sources at wavelengths shorter than $121\;{\rm nm}$. Using an attractive potential produced by a ${\rm CO}_2$ laser could lead to lifetimes well over a minute.

State of the art ${\rm CO}_2$ lasers can produce more than $10\;{\rm kW}$ of power. Using $25\;{\rm kW}$ of $10.6\;\mu{\rm m}$ light produced by a ${\rm CO}_2$ laser would give a shot-noise limit 100 times lower than the $50\;{\rm W}$ of $532\;{\rm nm}$ light discussed in the paper. Analyzing the impact of laser intensity noise in an attractive, harmonic trap is beyond the scope of this paper. However, even the repulsive optical potentials described in this paper could potentially have far lower $\delta P/P$ by using power build-up cavities to produce the trapping potentials.

Finally, at the price and size scale of present gravitational detectors, one could imagine a set of machines producing many BECs for simultaneous measurements to further improve sensitivity.

Though using a BEC to detect gravitational waves will be extremely difficult, it should be noted that LIGO was first conceptualized in the 1980s and only detected gravitational waves in 2015. BECs have a head start in knowing that gravitational waves {\it can} be directly detected and some of LIGO's technical innovations could potentially be adapted for use by a BEC GW detector. Though such an experiment may take years or decades to achieve, doing so will provide a new method to detecting gravitational waves across potentially several orders of magnitude in frequency.

\section*{Acknowledgements}

We thank Aditya Dhumuntarao, Soham Mukherjee, and Erickson Tjoa for their useful comments and discussions. We also thank the participants of the mini-symposium on Quantum Sensors and Their Applications in Fundamental Physics and Astrophysics Experiments (December 13, 2019) at Washington University in St. Louis for their helpful questions and suggestions. We thank Vladimir Dergachev for his valuable discussion regarding LIGO's calculations. We also thank Denis Martynov for providing LIGO's noise spectrum.

MR was funded by a National Science and Engineering Research Council of Canada (NSERC) graduate scholarship. AOJ was funded by by the University of Waterloo and the Institute for Quantum Computing. This research was supported in part by NSERC and the Perimeter Institute for Theoretical Physics, and by AOARD Grant FA2386-19-1-4077.
Research at Perimeter Institute is supported by the Government of Canada through Industry Canada and by the Province of Ontario through the Ministry of Economic Development \& Innovation. 

\bibliographystyle{unsrt}
\bibliography{Bibliography-MagneticFieldDecember2019}

\appendix

\section{Sensitivity to Gravitational Waves}
\label{sec: derivation}

As mentioned in Section \ref{sec: Estimation}, an unknown parameter in a quantum system can be estimated with the quantum Cramer-Rao bound, \cite{Braustein1994}
\begin{align}
\braket{(\Delta \epsilon)^2}\geq\frac{1}{MH_\epsilon}\ ,
\label{eq: Delta epsilon}
\end{align}
with $\braket{(\Delta \epsilon)^2}$ the expectation value of the uncertainty in $\epsilon$, $M$ the number of measurements of the system, and $H_\epsilon$ the Fisher information  \cite{Braustein1994}
\begin{align}
H_{\epsilon}=\frac{8\bigg(1-\sqrt{F(\rho_{\epsilon},\rho_{\epsilon+d\epsilon})}\bigg)}{d\epsilon^2}\ ,
\label{eq: QFIappendix}
\end{align}
where
 $F(\rho_{\epsilon},\rho_{\epsilon+d\epsilon})=\left[\Tr\sqrt{\rho_\epsilon\sqrt{\rho_{\epsilon+d\epsilon}}\rho_\epsilon}\right]^2$ is the fidelity \cite{Uhlmann1976,Jozsa1994}. Note that we can recast equations \eqref{eq: alpha rel} and \eqref{eq: beta rel} as $\alpha_{rel}=e^{-i\theta_{\alpha}}\cosh r_{rel}$ and $\beta_{rel}=e^{-i\theta_{\beta}}\cosh r_{rel}$, where $r_{rel}$ is the squeezing that results from the gravitational wave and $e^{-i\theta_{\alpha}},e^{-i\theta_{\beta}}$ are phase factors. We can then determine the $r_{rel}=\log\left[|\alpha_{rel}|+|\beta_{rel}|\right]$.
 
Let us suppose that before a gravitational wave interacts with the condensate, our phonons are in a squeezed state, with squeezing parameter $r_0$ and quadrature angle $\phi_0$, where the state of the phonons is $\ket{\zeta_0}=S_0\ket{0}=\exp\bigg[\frac{1}{2} (\zeta_0^* \hat a^2-\zeta_0 \hat a^{\dagger 2})\bigg]\ket{0}$,  $S_0$ is the squeezing operator, and $\zeta_0=r_0e^{i\phi_0}$ 

Let $\epsilon=\epsilon_1$ and $\epsilon_2=\epsilon+d\epsilon$. In a pure state, $H_{\epsilon}=\frac{8(1-|\braket{\epsilon_1|\epsilon_2}|)}{(\epsilon_1-\epsilon_2)^2}$. The state after interaction with a gravitational wave is $\ket{\epsilon_i}=S_{\epsilon_i}S_0\ket{0}$, where $S_{\epsilon_i}$ encodes the gravitational wave's influence. Then, $S_{rel}:=S_{\epsilon_1}^{\dagger}S_{\epsilon_2}$, we have 
\begin{align}
\braket{\epsilon_1|\epsilon_2}&=\braket{0|S_0^{\dagger}S_{rel}S_0|0}\\
&=\braket{0|S_0^{\dagger}\exp\bigg[\frac{1}{2} (\zeta_{rel}^* \hat a^2-\zeta_{rel} \hat a^{\dagger 2})\bigg]S_0|0}
\end{align}
where $\zeta_{rel}=r_{rel}e^{i\phi_{rel}}$. Noting that $r_{rel}\propto\epsilon$, we have
\begin{align}
    \braket{\epsilon_1|\epsilon_2}\sim1+\frac{1}{2}\braket{0|S_0^\dagger\left(\zeta_{rel}^* \hat a^2-\zeta_{rel} \hat a^{\dagger 2}\right) S_0|0}+\frac{1}{8}\braket{0|S_0^\dagger\left(\zeta_{rel}^* \hat a^2-\zeta_{rel} \hat a^{\dagger 2}\right)^2 S_0|0}
\end{align}
Thus, the quantum Fisher information is
\begin{align}
   (\epsilon_1-\epsilon_2)^2H_\epsilon=|\braket{0|S_0^\dagger\left(\zeta_{rel}^* \hat a^2-\zeta_{rel} \hat a^{\dagger 2}\right) S_0|0}|^2+\frac{1}{4}\braket{0|S_0^\dagger\left(\zeta_{rel}^* \hat a^2-\zeta_{rel} \hat a^{\dagger 2}\right)^2 S_0|0}
\end{align}
where we note that the first-order terms in $\epsilon$ vanish because of antisymmetry.
Using $S_0^{\dagger}aS_0=a\cosh r_0-a^{\dagger}e^{i\phi_0}\sinh r_0$ and $S_0^{\dagger}a^{\dagger}S_0=a^{\dagger}\cosh r_0-a e^{-i\phi_0}\sinh r_0$, we find

\begin{align}
     \braket{0|a^{\dagger2}|0}&=-e^{-i\phi_0}\cosh r_0\sinh r_0\\
     \braket{0|a^{2}|0}&=-e^{i\phi_0}\cosh r_0\sinh r_0\\
    \braket{0|a^{\dagger4}|0}&=3e^{-2i\phi_0}\cosh^2r_0\sinh^2r_0\\
    \braket{0|a^4|0}&=3e^{2i\phi_0}\cosh^2r_0\sinh^2r_0\\
    \braket{0|a^2a^{\dagger2}|0}&=2\cosh^4r_0+\cosh^2r_0\sinh^2r_0\\
    \braket{0|a^{\dagger2}a^2|0}&=2\sinh^4r_0+\cosh^2r_0\sinh^2r_0
\end{align}

Therefore,

\begin{equation}
\begin{aligned}
    |\braket{\epsilon_1|\epsilon_2}|\sim1&+\frac{r_{rel}^2}{2}\cosh^2r_0\sinh^2r_0\left[\sin^2(\phi-\phi_{rel})+3\cos(2\phi_0-2\phi_{rel})\right]\\
    &-\frac{r_{rel}^2}{2}\left(\cosh^4r_0+\sinh^4r_0+\cosh^2r_0\sinh^2r_0\right)
\end{aligned}
\end{equation}

Therefore, after averaging over angles, we find
\begin{align}
    (\epsilon_1-\epsilon_2)^2H_{\epsilon}=\frac{1}{4}r_{rel}^2\left(\cosh^4r_0+\sinh^4r_0\right)
    \label{eq: Hfinal}
\end{align}

Recalling that $r_{rel}=\log\left[|\alpha_{rel}|+|\beta_{rel}|\right]$ and from equations \eqref{eq: alpha rel}-\eqref{eq: beta rel},
\begin{align}\
\alpha_{rel}^*&=1+\tilde\epsilon\left(\alpha^*_{\tilde\epsilon}\alpha_{2a}-\beta_{\tilde\epsilon}^*\beta_{2a}\right)\ ,\\
\beta_{rel}^*&=-\tilde\epsilon\left(\alpha^*_{\tilde\epsilon}\beta^*_{2a}-\beta_{\tilde\epsilon}^*\alpha_{2a}^*\right)\ ,
\end{align}
 we find $r_{rel}\sim \left(\Re\left[\alpha_{rel}^{(1)}\right]+|\beta_{rel}|\right)^2\epsilon^2$, where $\alpha_{rel}=1+\epsilon\alpha_{rel}^{(1)}$ Therefore, with equations \eqref{eq: Delta epsilon} and \eqref{eq: Hfinal}, the sensitivity to gravitational waves is given by
 
 \begin{align}
 \braket{(\Delta\tilde\epsilon)^2}=\frac{4}{M\left(\Re\left[\alpha_{rel}^{(1)}\right]+|\beta_{rel}|\right)^2\left(\cosh^4r_0+\sinh^4r_0\right)}\ .
 \end{align}

\end{document}